\documentclass[fleqn,usenatbib]{mnras}

\usepackage{newtxtext,newtxmath}

\usepackage[T1]{fontenc}

\DeclareRobustCommand{\VAN}[3]{#2}
\let\VANthebibliography\thebibliography
\def\thebibliography{\DeclareRobustCommand{\VAN}[3]{##3}\VANthebibliography}

\usepackage{graphicx}	% Including figure files
\usepackage{amsmath}	% Advanced maths commands

\usepackage{graphicx}     % To include images
\usepackage{caption}      % For captioning
\usepackage{subcaption}   % For subfigures
\usepackage{pgfplots}     % For plotting
\usepackage{tikz}         % For custom annotations
\pgfplotsset{compat=1.17}
\usepackage{booktabs} % For professional looking tables
\usepackage{siunitx}  % For aligning numbers
\usepackage{siunitx}
\usepackage{rotating} % For tall table
\usepackage{makecell} % for multi-line headers

\title[Radio Imaging of PSR J1818-1607]{Low Frequency Radio Imaging Study of PSR J1818$-$1607}

\author[Yadav et al.]{
A. Yadav,$^{1}$\thanks{E-mail: abhinandan.frb@gmail.com}
M. P. Surnis,$^{1}$\thanks{E-mail: msurnis@iiserb.ac.in}
B. C. Joshi,$^{2,3}$
and M. Bagchi,$^{4,5}$\\
$^{1}$Department of Physics, IISER Bhopal, Bhauri Bypass Road, Bhopal, 462066, Madhya Pradesh, India\\
$^{2}$National Centre for Radio Astrophysics, SP Pune University Campus, Pune, 411007, Maharashtra, India\\
$^{3}$Department of Physics, Indian Institute of Technology Roorkee, Roorkee, 247667, Uttarakhand, India\\
$^{4}$The Institute of Mathematical Sciences, C.I.T. Campus, Taramani, Chennai, 600113, Tamilnadu, India\\
$^{5}$Homi Bhabha National Institute, Training School Complex, Anushakti Nagar, Mumbai, 400094, Maharashtra, India\\
}

\date{Accepted XXX. Received YYY; in original form ZZZ}

\pubyear{\the\year{}}

\begin{document}
\label{firstpage}
\pagerange{\pageref{firstpage}--\pageref{lastpage}}
\maketitle

% Abstract of the paper
\begin{abstract}
We report on the low-frequency radio observations of magnetar PSR J1818$-$1607 carried out with the upgraded Giant Metrewave Radio Telescope at band 3 (300$-$500 MHz) and band 4 (550$-$750 MHz). We have identified the continuum source associated with the magnetar and report variations in flux density and its spectral index. The flux density timeseries for the magnetar reveals the presence of two potential radio flaring episodes with varying spectral properties. We discuss the implications of the spectral index changes on the potential emission mechanisms for radio-loud magnetars. We also report non-detections of the continuum source as well as no direct indication for an associated diffuse emission from multiple archival radio imaging surveys.  
\end{abstract}

% Select between one and six entries from the list of approved keywords.
% Don't make up new ones.

%%%%%%%%%%%%%%%%%%%%%%%%%%%%%%%%%%%%%%%%%%%%%%%%%%

%%%%%%%%%%%%%%%%% BODY OF PAPER %%%%%%%%%%%%%%%%%%

\section{Introduction}

Magnetars are young neutron stars with very high magnetic fields ($10^{14}-10^{15}$ G), long spin periods ($1-12$ s), high spin-down rates ($10^{-13} - 10^{-11}$ ${\rm s~s^{-1}}$) and characteristic ages of a few hundred to a few thousand years \citep{KB17annurev}. They are typically discovered via their X-ray outbursts and are characterized by highly variable emission and irregularities in their rotation \citep{crh+06}. A normal radio pulsar (rotation powered) emits electromagnetic radiation by converting a small part of its spin-down energy \citep[e.g.][]{rs75}, while a magnetar emits via the dissipation of ultra-strong magnetic fields in the interior of the star \citep{dt92}.

There are a total of 30 magnetars (24 confirmed, 6 candidates) in our Galaxy out of which only five have shown radio pulsations, and one, SGR 1935+2154, has shown FRB-like burst \citep{ok14}\footnote{\url{http://www.physics.mcgill.ca/~pulsar/magnetar/main.html}}. Observational characteristics of the radio-loud magnetars reveal that their pulsed radio emission is more variable than those of normal pulsars in terms of flux density, profile shape, and polarization \citep[e.g.][]{crh+07, ksj+07, lbb+10, kjl+11, lek+13, sj13, lsj+20, ksj+21, ybm+19}. These emission characteristics are also shared with high magnetic field strength pulsars that occasionally display magnetar-like outbursts \citep{djw+18}. 

The radio flux density of radio-loud magnetars is one of their most distinctive and dynamic observational properties. Unlike rotation-powered pulsars, which generally exhibit relatively stable flux densities and steep radio spectra, magnetars show extreme and intrinsic variability, with their flux density sometimes changing by factors of a few on timescales ranging from minutes to days and evolving substantially over months to years\citep{crh+06, crh+07}. Their radio spectra are typically flat or inverted, with spectral indices of $\alpha \approx 0$ or $\alpha > 0$, in contrast to the steep negative spectra commonly observed in ordinary pulsars ($\alpha \sim -1.8$) \citep{crj+08, lbb+10, ybm+19}. Consequently, magnetars can remain unusually bright at high radio frequencies, with emission detected at frequencies extending up to $\sim 353$ GHz \citep{tbb+22}. The radio flux density is often strongly influenced by the magnetospheric state of the neutron star and may increase dramatically following an X-ray outburst before gradually declining as the source recovers, although an episode of radio rebrightening without a corresponding X-ray burst was also observed for XTE J1810$-$197 \citep{lls+18, mnk+18}. The same magnetar also exhibited Gigahertz-Peaked Spectra (GPS) in which, the flux density reaches a maximum at a characteristic frequency that was observed to shift towards lower frequencies during the post-outburst recovery phase \citep{msj+22}. These diverse spectral and temporal variations make flux-density monitoring across a broad frequency range an important tool for probing the evolving magnetospheric conditions of magnetars. Radio observations generally measure period-averaged and peak flux densities, while rare, extremely bright individual bursts can be characterized in terms of their fluence; the most remarkable example is SGR 1935$+$2154, which produced bright millisecond-duration radio bursts with fluences reaching the MJy ms level, providing a crucial connection between magnetar radio emission and the broader phenomenon of fast radio bursts (FRBs) \citep{ksj+21}.

PSR J1818$-$1607 was discovered via an X-ray outburst on 12 March 2020 by the Burst Alert Telescope (BAT) onboard the Neil Gehrels Swift X-ray observatory \citep{egk+20}. Data from the Neutron Star Interior Composition Explorer (NICER) revealed a coherent periodicity of 0.733417(4) Hz (or a rotation period of $\sim1.36$ s). Soon after the X-ray outburst, a series of follow-up radio observations started, detecting radio pulsations with a period of 0.737410(2) Hz by the 100 m Effelsberg telescope \citep{lkc+20}; later confirmed by the Lovell telescope \citep{rsl+20} and at the low frequency by the upgraded Giant Metrewave Radio Telescope (uGMRT) \citep{jb20}. Radio emission from this magnetar has a very high degree (80 to 90\%) of linear polarization. The spin period derivative of $\dot{P} = 9 \pm 1 \times 10^{-11}$ ${\rm s~s^{-1}}$ reported from radio timing measurements \citep{rsl+22} resulted in the first estimate of the surface dipolar magnetic field strength of $B = 3.4 \times 10^{14}$ G and the spin-down luminosity of $\dot{E} = 1.3 \times 10^{36}$ erg s$^{-1}$, which confirmed the magnetar nature of the source. PSR J1818-1607 exhibits highly variable radio emission with a strong dependence on observing frequency. Its radio spectrum evolved from an initially steep spectrum to a much flatter one, and later observations revealed a GPS with a peak frequency of $5.4 \pm 0.6$ GHz, followed by a steep decline in flux density at higher frequencies \citep{lbl+25}.

In a recent paper, \cite{ibr+23} used observations carried out with the Jansky Very Large Array (JVLA) at 3 GHz to propose that a diffuse radio shell-like emission of size $90''$ was present around the magnetar. If such a supernova remnant (SNR) were to be confirmed and firmly associated with the magnetar, it would imply that PSR J1818$-$1607 is much older than what was previously suggested. It is thus very important to search for the presence of such an SNR through imaging observations at other radio frequencies.
  
In this paper, we report the results of radio imaging observations carried out using the uGMRT \citep{gak+17}. The organization of the paper is as follows: Section \ref{sec:obs} describes the details of observations and analysis of the data is presented in Section \ref{sec:data}. Section \ref{sec:res} deals with results and their implications. In Section \ref{sec:conc} we conclude our findings.        

\section{Observations}
\label{sec:obs}

We observed PSR J1818$-$1607 with the uGMRT on a total of 19 epochs using the band 4 (550$-$750 MHz) and 16 epochs using the band 3 (300$-$500 MHz) receivers. The observations spanned the date range MJD 59554--60373 (2021 June 12 -- 2024 April 3) with a monthly cadence. The band 3 and the band 4 observations were carried out either the same day or within a day from each other. The beamformed data at both bands were recorded in the phased array (PA) mode over 4096 frequency channels and sampled every 1.3 milliseconds. The PA consisted of the 14 central square antennae and one antenna from each of the three arms of the uGMRT. The interferometry data were recorded with a default sampling time of 10.7 seconds over the same number of channels. We observed 3C286 as the flux density and bandpass calibrator and 1822$-$096 as the phase calibrator at each epoch. We also observed PSR B1822$-$09 as a test pulsar for five minutes at each epoch to make sure that the settings were working as expected for the beamformer mode.  

In addition to this, we also used the interferometry data from uGMRT band 3 and band 4, just after the outburst (proposal codes: ddtC128 and ddtC129, PI: B.C. Joshi). These observations were carried out in the total intensity mode with 4096 channels and an integration time of 2.0132 sec. 

\section{Data Analysis}
\label{sec:data}

The present paper focuses on the results from the interferometry data. The analysis of the beamformer data is underway and the results from that analysis will be reported elsewhere. We thus describe the analysis of the interferometry data in this section. 

\subsection{Making Radio Images}
We used an automated imaging pipeline\footnote{\url{https://github.com/Kaustubh-Rai-28/uGMRT-PIPELINE}} that is based on an early version of the CAPTURE pipeline\footnote{\url{https://github.com/ruta-k/CAPTURE-CASA6}} \citep{ki+21}. After importing the data in the standard measurement set (MS) format, the first step is to flag bad data using the \texttt{flagdata} task in multiple rounds beginning with the 'quack' mode (flagging the first time slice data) followed by flagging very high and very low values. Flagged data are then calibrated. This is followed by another round of detailed flagging and a second round of calibration. The data are then imaged using the 'wprojection' and 'mtmfs' techniques in Common Astronomy Software Application\citep[CASA;][version~6.2]{casa22} using the task \texttt{tclean}. We used Briggs weighting with a robust parameter of 0 to achieve a balance between sensitivity and angular resolution. Additionally, images were also produced with robust parameters of 0.5, 1, and 2 in order to enhance the sensitivity to any potential diffuse emission. The cell size used for both the frequency bands was 1'' $\times$ 1''. The image obtained after the deconvolution was further improved by phase only self-calibration. The timescale for calculating phase solutions was 8, 4, 2, and 1 minute respectively, for the four iterations.

In addition, we also used imaging data after the last round of self-calibration from multiple epochs to produce a combined radio image through the task \texttt{tclean} with the same set of parameters as those used for imaging individual epochs. These images provided better sensitivity compared to individual epochs due to increased baseline coverage of the combined data set. 

In total, imaging was performed for all the epochs. To ensure data quality, we applied a $50\%$ threshold for flagged data. Epochs with more than $50\%$ flagged data were excluded from the composite image. Following this criterion, 6 epochs in band 3 and 11 epochs in band 4 were retained for further analysis. In this way, the signal-to-noise ratio (S/N) will increase by a factor of $\sqrt{N}$, where N is the number of epochs used for the combination. 

\subsection{Flux Density Measurement}
We measured the flux density of the magnetar in the Stokes~I images using the \texttt{imfit} task in CASA. We defined a square box of 20$\times$20 pixels centred on the position of the magnetar and fit for an elliptical Gaussian plus a constant background assuming an unresolved source. \texttt{imfit} returns the best-fit integrated flux density in Jy and the elliptical Gaussian widths (FWHM major/minor axes $\theta_{\rm maj}$, $\theta_{\rm min}$, and position angle). Uncertainties in the flux density were estimated by adding the image thermal noise and the absolute flux-scale error in quadrature. We then included a fractional absolute calibration uncertainty (typically 5$-$10\% for our band) to obtain more realistic results.

\subsection{Image Analysis}
\label{subsec:image_ana}
We selected five different source-free regions around the magnetar. We noted the root mean square (RMS) for these regions using the CARTA – Cube Analysis and Rendering Tool for Astronomy \citep{carta} statistics widget. We added the individual RMS values in quadrature to get the final RMS. This RMS per beam was used to put upper limits on any extended emission that was not detected.

\section{Results and Discussion}
\label{sec:res}

\subsection{Flux Density Evolution}
\begin{figure*}
     \centering
     \begin{subfigure}[t]{0.48\textwidth}
         \centering
         \includegraphics[width=0.99\linewidth]{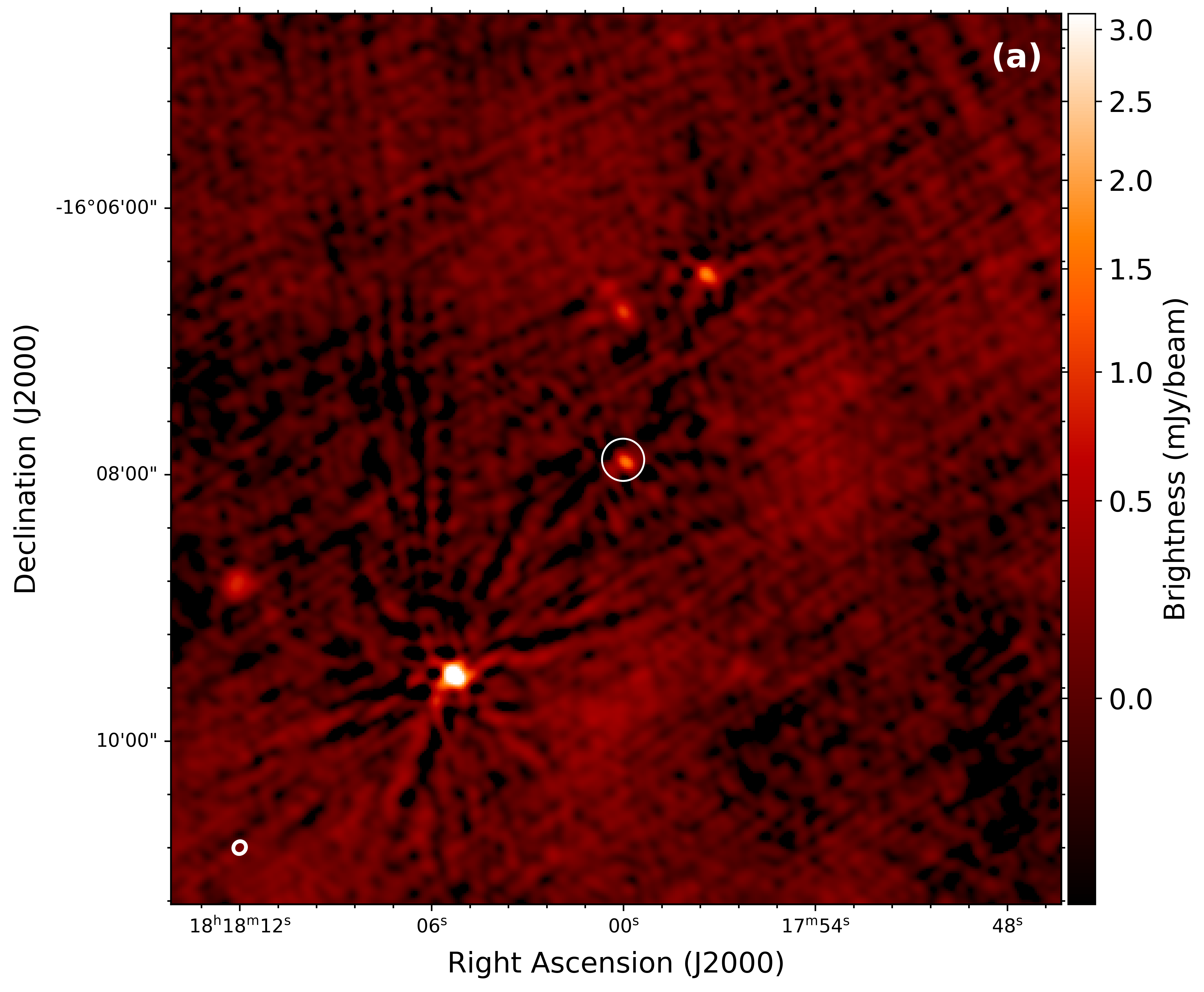} 
         %\caption{uGMRT image of PSR J1818-1607 at 407 MHz.}
         \label{fig:img_band3}
     \end{subfigure}
     \hfill
     \begin{subfigure}[t]{0.48\textwidth}
         \centering
         \includegraphics[width=\linewidth]{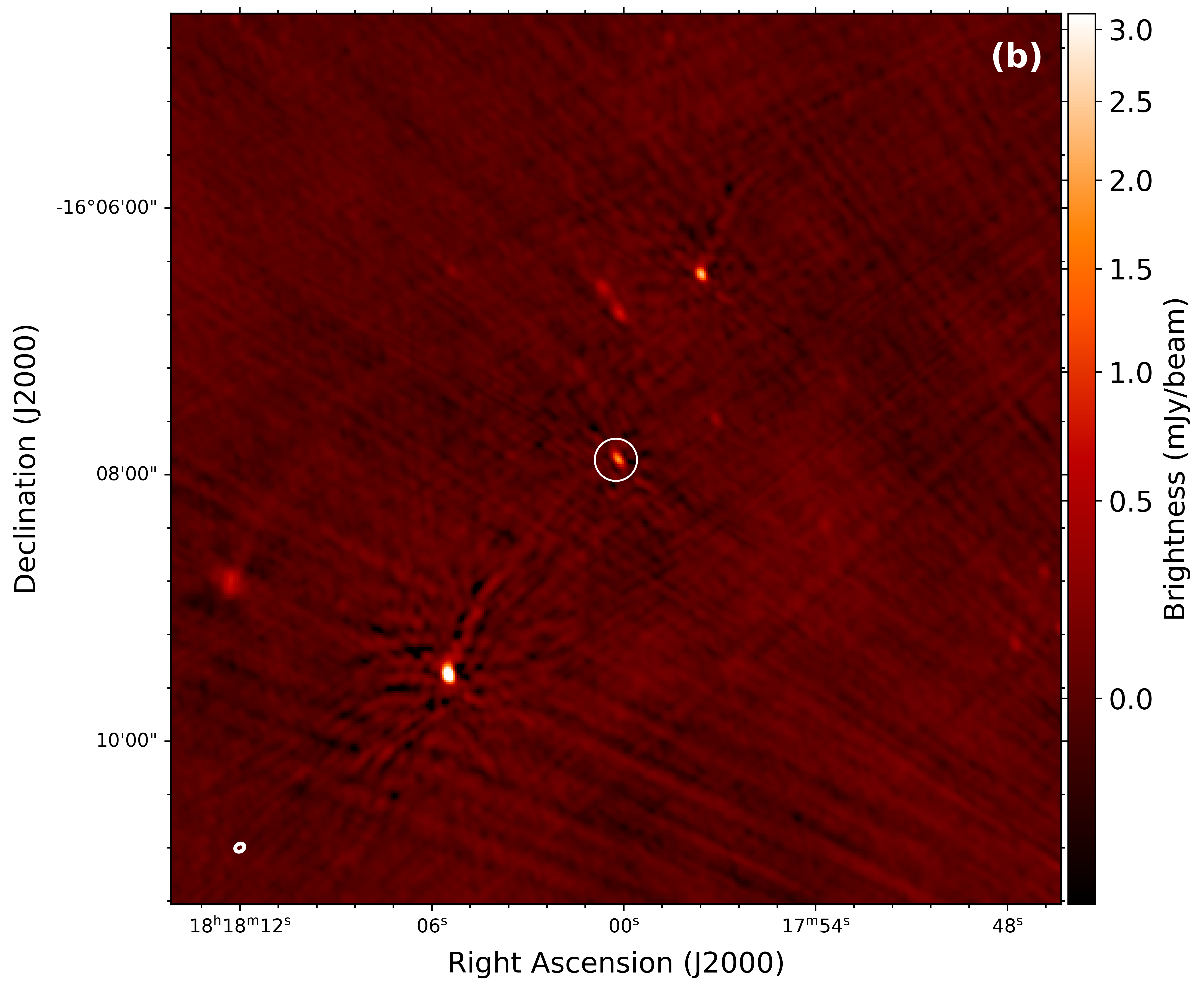}
         %\caption{uGMRT image of PSR J1818-1607 at 643 MHz.}
         \label{fig:img_band4}
     \end{subfigure}
    
     \caption{The uGMRT image of PSR J1818$-$1607 at 407 MHz (band 3; left panel) and 643 MHz (band 4; right panel). The continuum source corresponding to the magnetar is denoted by a white circle. The white ellipse at the bottom left corner shows the synthesised beam size. The colour scales of both images are similar and the difference in the contrast is mostly due to the background sky temperature being higher at 407 MHz compared to 643 MHz. The band 3 image is a combined image from data over 6 epochs, and the band 4 image is a combined image from data over 11 epochs. The image RMS is 3.14 mJy beam$^{-1}$ at 643 MHz and 10.68 mJy beam$^{-1}$ at 407 MHz, respectively.}
     \label{fig:combined}
 \end{figure*}

\begin{figure}
    \centering
    \includegraphics[width=\linewidth]{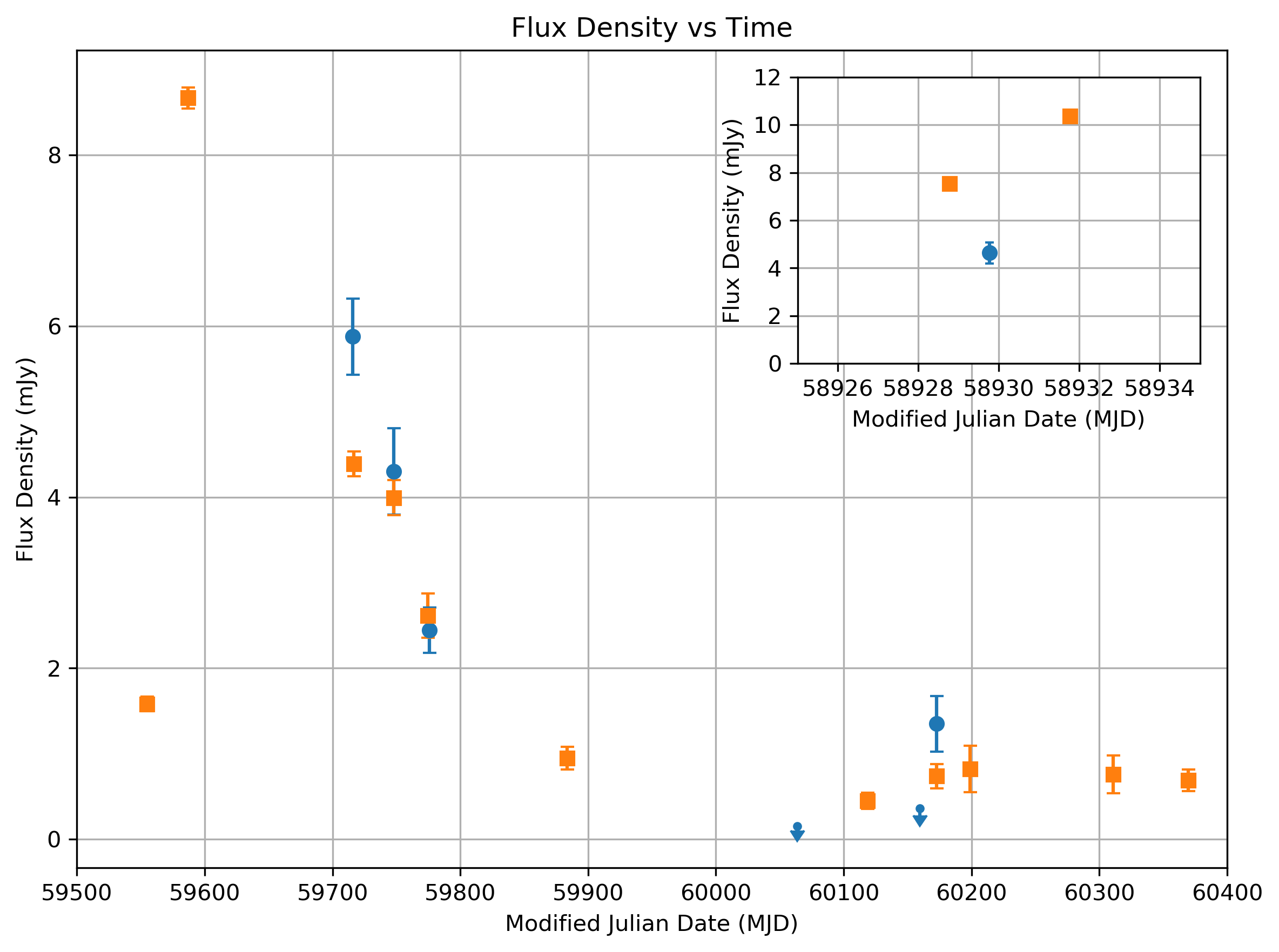}
    \caption{The flux density of PSR J1818-1607 as a function of the MJD. The teal circles represent the measurements at band 3, while the orange squares represent the measurements at band 4. The three points in the inset are from the observations following the initial outburst in March 2020.}
    \label{fig:flux_evo}
\end{figure}

We show the composite images from our uGMRT observations in Figure \ref{fig:combined}. The magnetar can be seen as a continuum source (inside the white circle) in the images at both frequencies. The immediate neighbourhood of the magnetar does not show any special extended or diffuse structure associated with it, as expected for an associated SNR shell. The overall flux density we estimate from these images is $1.9 \pm 0.3$ mJy at 407 MHz and $2.3 \pm 0.1$ at 643 MHz. We have also calculated the continuum flux density at each epoch for band 3 and band 4. Figure ~\ref{fig:flux_evo} shows the flux density time series for the magnetar. The first three flux densities (inset in Figure \ref{fig:flux_evo}) are just after the outburst in band 3 and band 4. We started the regular follow-up observations of the magnetar a year after the initial outburst stage. In our observations, we can clearly see that the magnetar has undergone another radio flare. As the flare decays, we see that the spectral index changes very rapidly. Initially, it shows a positive spectral index (brighter at band 4 than band 3) and then becomes flat, which is consistent with typical radio-loud magnetars. We see another mini-flare lasting roughly 300 days (or maybe more) after MJD 60100 (2023 June 5). Here, we see that the magnetar is not detected in band 3 for a couple of epochs but is detected at band 4 (indicating a positive spectral index). At the peak of this mini-flare, we see that the spectrum is typical of a radio pulsar with a negative spectral index. This is assuming that the non-detections in the band 3 observations on either side of MJD 60100 indicate the general trend for the flux density of the magnetar at band 3 frequencies.  

\subsection{Putative SNR Shell} 
\begin{figure}
    \centering
    \includegraphics[width=1\linewidth]{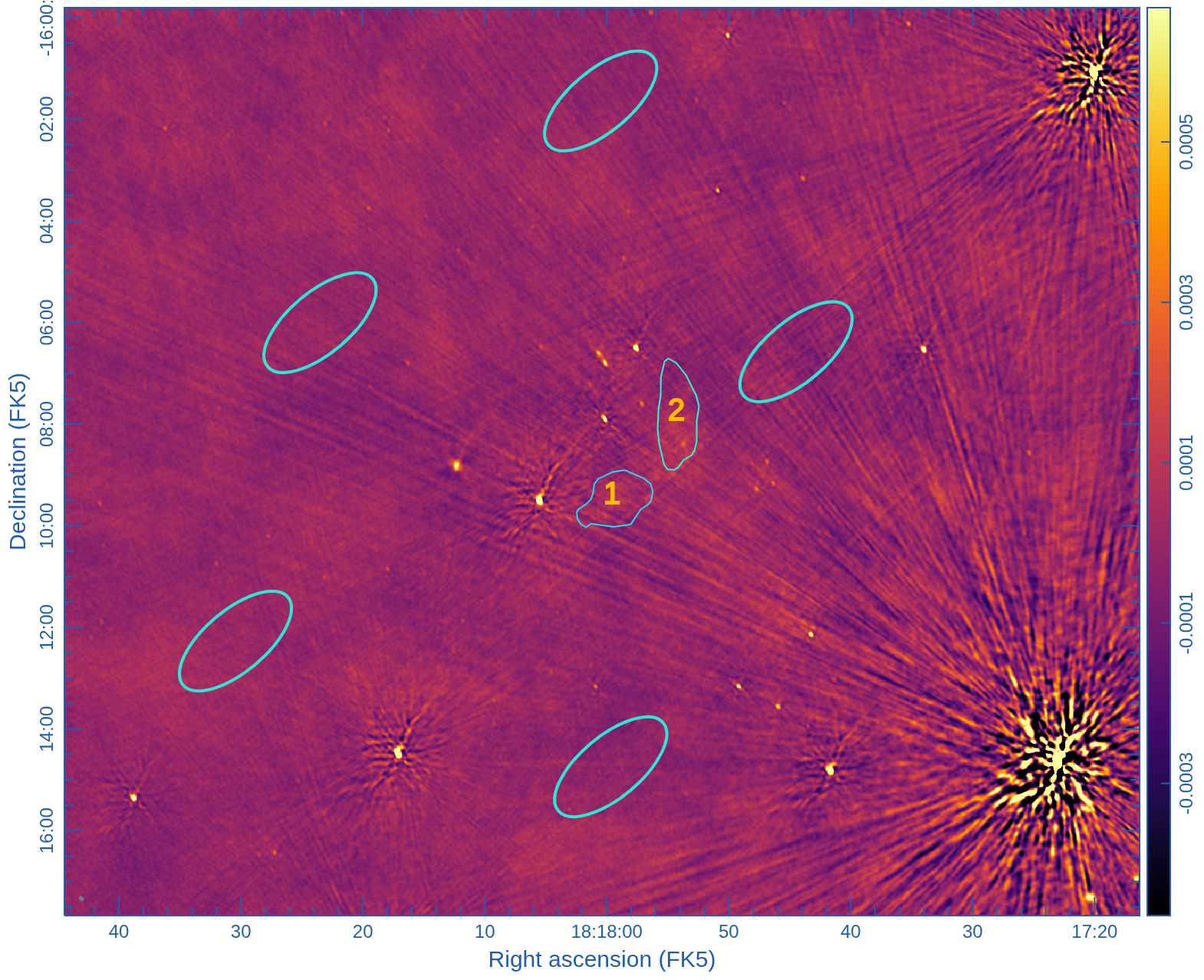}
    \caption{The band 4 image of the region with brightest contours from \protect\cite{ibr+23} plotted as two areas numbered 1 and 2. Additionally, there are five elliptical regions containing no sources. These regions were used to determine the statistics of the image in order to place upper limits on the significance of the non-detection of the putative SNR shell.}
    \label{fig:rms}
\end{figure}

One look at the composite images from our uGMRT observations (Figure \ref{fig:combined}) indicate that the putative SNR seen at 3 GHz with the JVLA may not be present. To place an upper limit on the significance of the SNR non-detection, we used the brightest contours from the radio image obtained by \cite{ibr+23} and overlaid them on our composite band 4 image, as shown in Figure~\ref{fig:rms}. We then multiplied square root of the total number of pixels enclosed within these contours by the RMS obtained from the five regions as explained in Section \ref{subsec:image_ana}. The $3\sigma$ upper limit was taken as three times the RMS thus obtained. In order to check for the presence of the putative SNR shell (as well as the magnetar itself), we searched the archival radio data at different frequencies prior to the magnetar outburst. The results of this archival data search are presented here and summarised in Table~\ref{tab:radio_observations}. We have provided brief details of the individual surveys below.

\subsubsection{TIFR GMRT Sky Survey (TGSS)}
 TGSS is a 150 MHz continuum survey by the Giant Metrewave Radio Telescope at a central frequency of 147.5 MHz and a bandwidth of 16.7 MHz. The particular field of interest was observed on 15 March 2016 and has an integration time of 15 min \citep{ijm+17}. We did not see the magnetar in the radio image and there was no direct indication of the putative SNR shell as well.
 
\subsubsection{Rapid ASKAP Continuum Survey (RACS)}
RACS is a Southern sky survey with the Australian Square Kilometre Array Pathfinder (ASKAP) in the frequency range of 700$-$1800 MHz. Depending on the ASKAP observing bands RACS is further divided into three surveys, namely RACS-low (887 MHz), RACS-mid (1367 MHz) and RACS-high (1655 MHz). RACS-low, mid and high have bandwidths of 288 MHz, 144 MHz and 200 MHz, respectively and integration time of 15 minutes per tile for all bands \citep{hmt+21,dgh+24}. We did not see the magnetar in the radio image and there was no direct indication of the putative SNR shell as well.

\subsubsection{SARAO MeerKAT Galactic Plane Survey (SMGPS)}
SMGPS is a 1.3 GHz continuum imaging survey by the South African Radio Astronomy Observatory (SARAO) Meer (More) Karoo Array Telescope (MeerKAT) covering a frequency range 856$-$1712 MHz. The observation were carried out between 2018$-$2020 and the particular field containing the magnetar was observed on 26 August 2018 \citep{SWF+24}. We did not see the magnetar in the radio image and there was no direct indication of the putative SNR shell as well.

\subsubsection{NRAO Very Large Array Sky Survey (NVSS)}
The NRAO VLA Sky Survey (NVSS) was conducted using the Very Large Array (VLA) at a frequency of 1.4\,GHz (L-band). The survey covers approximately 82\% of the sky, including all regions north of $-40^\circ$ declination. It provides images with an angular resolution of $\sim 45''$ and a typical rms sensitivity of $\sim 0.45$\,mJy\,beam$^{-1}$. The particular field of interest containing the magnetar (RA, Dec) was observed on 1996 June 19, and its RMS around the magnetar is 2635.87 mJy beam$^{-1}$. The high RMS region around the magnetar is likely caused by its proximity to a complex radio source. The complex source was imaged separately and later stitched into the original image, which may have introduced enhanced noise and imaging artefacts in the surrounding region. We did not see the magnetar in the radio image and there was no direct indication of the putative SNR shell as well.

\subsection{Discussion}

\subsubsection{Flux Density and Spectral Index Variations}

Our follow-up observations reveal two distinct flaring episodes occurring approximately one year after the initial X-ray outburst (see Figure \ref{fig:flux_evo}). Archival uGMRT observations obtained shortly after the outburst indicate that the PSR~J1818$-$1607 was already in radio-bright state at that time. The flux density seems to have decayed to a quiescent level just prior to the commencement of our monitoring campaign. The detection of renewed flaring activity therefore suggests that the post-outburst evolution is not characterized by a simple monotonic decay, but instead involves episodic rebrightening of the radio emission.

The measured peak flux densities during the flaring episodes exceed the quiescent level by more than a factor of four, indicating substantial variability in the magnetospheric emission properties. Such variability is consistent with the behaviour observed in other transient radio-loud magnetars, where large flux variations are frequently associated with magnetospheric reconfiguration following an outburst.

Magnetars in the immediate post-outburst phase often exhibit flat or inverted radio spectra ($\alpha \gtrsim 0$), in contrast to the steep negative spectra of ordinary rotation-powered pulsars. As the outburst decays, the spectral index may evolve toward flatter or even negative values in some cases, although this behaviour is not universal and can vary significantly between sources (e.g., XTE~J1810$-$197; \citealt{crh+06, msj+22}; PSR~J1622$-$4950; \citealt{lbb+10, scholz2017}).

In our observations, we find that during flaring episodes, the spectral index departs significantly from the quiescent value and becomes flat or positive. This behaviour suggests a change in the broadband emission properties associated with the flaring state. The spectral inversion indicates enhanced high frequency emission relative to low frequencies, implying a modification in the particle energy distribution and/or radiative transfer conditions within the magnetosphere.

The presence of multiple flaring episodes may be characteristic of recently activated or young magnetars. For example, PSR~J1622$-$4950 exhibited multiple episodes of radio flaring in archival observations \citep{scholz2017}. Recurrent flaring may therefore reflect ongoing magnetospheric instability or delayed relaxation following large-scale magnetic field rearrangement.

A useful comparison can be made with XTE~J1810$-$197, in which the spectral index evolved from positive to flat and subsequently to negative values over a timescale of $\sim 250$ days during its post-outburst evolution \citep{msj+22}. This indicates a gradual spectral softening as the magnetosphere relaxed toward a more stable configuration. In contrast, PSR~J1818$-$1607 exhibits spectral flattening or inversion specifically during discrete flaring episodes, rather than a smooth long term monotonic trend.

Another relevant case is PSR~J1119$-$6127, a high magnetic field radio pulsar that underwent a magnetar-like X-ray outburst on 2016 July 26. Following the outburst, the radio emission temporarily disappeared and subsequently reactivated. After reactivation, the flux density increased by approximately an order of magnitude compared to the pre-outburst level. Simultaneous dual-frequency observations reported by \citet{mpd+2017} showed that the spectral index evolved from $\alpha = -0.8$ prior to the outburst to $\alpha = -2.2(2)$ shortly after reactivation, and then to $\alpha = -1.9(2)$ ten days later. This evolution indicates spectral flattening during the decay phase, consistent with magnetospheric restructuring.

The observed flux density variability and spectral index evolution in PSR~J1818$-$1607 suggests that the radio emission is highly sensitive to changes in the magnetospheric state. The transition from steep to flat or inverted spectra during flaring episodes may indicate of variations in particle acceleration efficiency, changes in emission altitude, or modifications to the optical depth within the closed field line region. Continued multi-frequency monitoring of radio-loud magnetars is essential to characterize the temporal coupling between flux density enhancement and spectral evolution.

\subsubsection{Putative SNR and the Age of the Magnetar}
We did not see any diffuse emission around the magnetar either in our own radio images or the archival images where we specifically searched. \cite{ibr+23} reported shell-like diffuse radio emission around the magnetar peaking at $\sim 2.2$ mJy beam$^{-1}$ at 3 GHz. If we assume that the spectral index of SNR emission is $-$0.5, then the peak flux density of this diffuse emission would be around $\sim 4.8$ mJy beam$^{-1}$ at 643 MHz and $\sim 6$ mJy beam$^{-1}$ at 407 MHz, respectively. Since the image RMS in the most sensitive single epoch images from our observations is 0.03 mJy beam$^{-1}$ at 643 MHz and 0.12 mJy beam$^{-1}$ at 407 MHz, we should have clearly detected the peak of the diffusion emission in our radio images (see Appendix A for more detailed calculations regarding the upper limits on the diffuse emission). This, in addition to the lack of any diffuse emission from the archival radio images, points towards the possibility that the putative SNR may not be present. Looking at the images with weights favouring shorter baselines (Figures B1 through B4 in the appendix), it is clear that the only images with some hint of shell-like emission comes from the images with the robust value of 2 at both the bands (right panels of Figures B2 and B4). This indicates the possibility that this is most probably some spurious combination of background emission and the lack of enough short baselines in the dataset. To that extent, the SMGPS survey image provides a decent upper limit. The expected peak value at 1300 MHz with a spectral index of $-$0.5 is about 3.3 mJy beam$^{-1}$ which is slightly above the 3$\sigma$ limit from Table \ref{tab:radio_observations}. Additionally, the candidate SNR catalog from the SMGPS \citep{acf+25} does not list any SNR candidate in the vicinity of the magnetar. MeerKAT has a much better short baseline coverage compared to the uGMRT (and hence, a better surface brightness sensitivity), and the lack of shell-like emission in the SMGPS image provides a slightly more constraining argument that the SNR may not be present. SNRs typically evolve through four stages: the free expansion phase, the Sedov-Taylor phase, the radiative phase, and the dispersion phase. The SNR shell becomes prominently visible during the Sedov-Taylor phase, which typically begins after a few thousand years \citep[e.g.][]{blr01}. The non-detection of any associated SNR shell with deeper future observations will help put an upper bound on the potential age of the magnetar. Vice versa, the detection of an SNR shell will definitively put a lower bound of a few thousand years on the age of PSR J1818$-$1607.    
  
\begin{table*}
    \centering
    \caption{$3\sigma$ upper limits on the diffuse emission around PSR~J1818$-$1607 at different observing frequencies from various radio imaging surveys.}
    \label{tab:radio_observations}
    \setlength{\tabcolsep}{10pt} % Adjust column spacing if necessary
    \begin{tabular}{lS[table-format=1.2]S[table-format=3.2]l} 
        \toprule
        \textbf{Telescope} & \textbf{Frequency} & \textbf{3$\sigma$ upper limit} & \textbf{Reference} \\
                      & \textbf{(MHz)}  & \textbf{(mJy beam$^{-1}$)} & \\ 
        \midrule
        TGSS & 150  & 2814.12 & \cite{ijm+17} \\
        uGMRT & 407   & 33.00 & This work \\
        uGMRT & 643   & 9.72 & This work \\
        % RACS-Low& 887   & 375.48  & \cite{hmt+21} \\
        RACS-Mid& 1367  & 122.28 & \cite{dgh+24} \\
        RACS-High & 1655  & 145.08 & \cite{dgh+24} \\
        SMGPS & 1300 & 254.40 & \cite{SWF+24} \\
        % NVSS & 1400 & 2635.87 & \cite{NVSS} \\
        \bottomrule
    \end{tabular}
\end{table*}

\section{Conclusions}
\label{sec:conc}

We have studied the magnetar PSR J1818-1607 with the uGMRT in band 3 and band 4. We have detected the magnetar as a continuum source at both frequencies, and the flux density from stacked images is $1.9 \pm 0.3$ mJy at 407 MHz and $2.3 \pm 0.1$ at 643 MHz, confirming the flat spectrum at radio wavelengths. We see no direct indication for the presence of the putative shell-like SNR in the vicinity of the magnetar. This implies that the magnetar is probably on the younger side. Thus putting it with the typical radio-loud magnetar population, which is a younger class of neutron stars.

\section*{Acknowledgements}
AY and MPS acknowledge the use of computational infrastructure built through the start-up grant awarded by IISER Bhopal. They would also like to thank Dr. Yogesh Maan for fruitful discussions. We acknowledge the support of the uGMRT staff in resolving technical issues and telescope operators for the observations. The uGMRT is run by the National Centre for Radio Astrophysics of the Tata Institute of Fundamental Research, India. The MeerKAT telescope is operated by the South African Radio Astronomy Observatory, which is a facility of the National Research Foundation, an agency of the Department of Science and Innovation.

%%%%%%%%%%%%%%%%%%%%%%%%%%%%%%%%%%%%%%%%%%%%%%%%%%
\section*{Data Availability}

The data used in this manuscript are publicly available from the GMRT online archive at \url{https://naps.ncra.tifr.res.in/goa/data/search}. Please use proposal codes 41\_059, 42\_041, 43\_028, 44\_003 and 45\_008.

%%%%%%%%%%%%%%%%%%%% REFERENCES %%%%%%%%%%%%%%%%%%

% The best way to enter references is to use BibTeX:

\bibliographystyle{mnras}
\bibliography{references} % if your bibtex file is called example.bib

@article{KB17annurev,
   author = "Kaspi, Victoria M. and Beloborodov, Andrei M.",
   title = "Magnetars", 
   journal= "Annual Review of Astronomy and Astrophysics",
   year = "2017",
   volume = "55",
   number = "Volume 55, 2017",
   pages = "261-301",
   doi = "https://doi.org/10.1146/annurev-astro-081915-023329",
   url = "https://www.annualreviews.org/content/journals/10.1146/annurev-astro-081915-023329",
   publisher = "Annual Reviews",
   issn = "1545-4282",
   type = "Journal Article",
  }

@ARTICLE{crh+07,
       author = {{Camilo}, F. and {Ransom}, S.~M. and {Halpern}, J.~P. and {Reynolds}, J.},
        title = "{1E 1547.0-5408: A Radio-emitting Magnetar with a Rotation Period of 2 Seconds}",
      journal = {\apjl},
         year = 2007,
        month = sep,
       volume = {666},
       number = {2},
        pages = {L93-L96},
          doi = {10.1086/521826},
archivePrefix = {arXiv},
       eprint = {0708.0002},
 primaryClass = {astro-ph},
       adsurl = {https://ui.adsabs.harvard.edu/abs/2007ApJ...666L..93C}
}

@ARTICLE{dt92,
       author = {{Duncan}, Robert C. and {Thompson}, Christopher},
        title = "{Formation of Very Strongly Magnetized Neutron Stars: Implications for Gamma-Ray Bursts}",
      journal = {\apjl},
         year = 1992,
        month = jun,
       volume = {392},
        pages = {L9},
          doi = {10.1086/186413},
       adsurl = {https://ui.adsabs.harvard.edu/abs/1992ApJ...392L...9D}
}

@ARTICLE{ok14,
       author = {{Olausen}, S.~A. and {Kaspi}, V.~M.},
        title = "{The McGill Magnetar Catalog}",
      journal = {\apjs},
         year = 2014,
        month = may,
       volume = {212},
       number = {1},
          eid = {6},
        pages = {6},
          doi = {10.1088/0067-0049/212/1/6},
archivePrefix = {arXiv},
       eprint = {1309.4167},
 primaryClass = {astro-ph.HE},
       adsurl = {https://ui.adsabs.harvard.edu/abs/2014ApJS..212....6O}
}

@ARTICLE{crh+06,
       author = {{Camilo}, Fernando and {Ransom}, Scott M. and {Halpern}, Jules P. and {Reynolds}, John and {Helfand}, David J. and {Zimmerman}, Neil and {Sarkissian}, John},
        title = "{Transient pulsed radio emission from a magnetar}",
      journal = {\nat},
         year = 2006,
        month = aug,
       volume = {442},
       number = {7105},
        pages = {892-895},
          doi = {10.1038/nature04986},
archivePrefix = {arXiv},
       eprint = {astro-ph/0605429},
 primaryClass = {astro-ph},
       adsurl = {https://ui.adsabs.harvard.edu/abs/2006Natur.442..892C}
}

@ARTICLE{ksj+07,
       author = {{Kramer}, M. and {Stappers}, B.~W. and {Jessner}, A. and {Lyne}, A.~G. and {Jordan}, C.~A.},
        title = "{Polarized radio emission from a magnetar}",
      journal = {\mnras},
         year = 2007,
        month = may,
       volume = {377},
       number = {1},
        pages = {107-119},
          doi = {10.1111/j.1365-2966.2007.11622.x},
archivePrefix = {arXiv},
       eprint = {astro-ph/0702365},
 primaryClass = {astro-ph},
       adsurl = {https://ui.adsabs.harvard.edu/abs/2007MNRAS.377..107K}
}

@ARTICLE{lbb+10,
       author = {{Levin}, Lina and {Bailes}, Matthew and {Bates}, Samuel and {Bhat}, N.~D. Ramesh and {Burgay}, Marta and {Burke-Spolaor}, Sarah and {D'Amico}, Nichi and {Johnston}, Simon and {Keith}, Michael and {Kramer}, Michael and {Milia}, Sabrina and {Possenti}, Andrea and {Rea}, Nanda and {Stappers}, Ben and {van Straten}, Willem},
        title = "{A Radio-loud Magnetar in X-ray Quiescence}",
      journal = {\apjl},
         year = 2010,
        month = sep,
       volume = {721},
       number = {1},
        pages = {L33-L37},
          doi = {10.1088/2041-8205/721/1/L33},
archivePrefix = {arXiv},
       eprint = {1007.1052},
 primaryClass = {astro-ph.HE},
       adsurl = {https://ui.adsabs.harvard.edu/abs/2010ApJ...721L..33L}
}

@ARTICLE{kjl+11,
       author = {{Keith}, M.~J. and {Johnston}, S. and {Levin}, L. and {Bailes}, M.},
        title = "{17- and 24-GHz observations of southern pulsars}",
      journal = {\mnras},
         year = 2011,
        month = sep,
       volume = {416},
       number = {1},
        pages = {346-354},
          doi = {10.1111/j.1365-2966.2011.19041.x},
archivePrefix = {arXiv},
       eprint = {1105.3961},
 primaryClass = {astro-ph.SR},
       adsurl = {https://ui.adsabs.harvard.edu/abs/2011MNRAS.416..346K}
}

@ARTICLE{lek+13,
       author = {{Lee}, K.~J. and {Eatough}, Ralph and {Karuppusamy}, Ramesh and {Champion}, David and {Keane}, Evan and {Kramer}, Michael and {Schnitzeler}, Dominic and {Noutsos}, Aris and {Klein}, Bernd and {Kraus}, Alex and {Bassa}, Cees and {Lyne}, Andrew and {Stappers}, Ben and {Spitler}, Laura and {Freire}, Paulo and {Cognard}, Ismael and {Desvignes}, Gregory and {Lazarus}, Patrick and {Verbiest}, Joris and {Brunthaler}, Andreas and {Falcke}, Heino},
        title = "{Polarisation profiles and rotation measure of PSR J1745-2900 measured at Effelsberg}",
      journal = {The Astronomer's Telegram},
         year = 2013,
        month = may,
       volume = {5064},
        pages = {1},
       adsurl = {https://ui.adsabs.harvard.edu/abs/2013ATel.5064....1L}
}

@ARTICLE{sj13,
       author = {{Shannon}, R.~M. and {Johnston}, S.},
        title = "{Radio properties of the magnetar near Sagittarius a* from observations  with the australia telescope compact array.}",
      journal = {\mnras},
         year = 2013,
        month = aug,
       volume = {435},
        pages = {L29-L32},
          doi = {10.1093/mnrasl/slt088},
archivePrefix = {arXiv},
       eprint = {1305.3036},
 primaryClass = {astro-ph.HE},
       adsurl = {https://ui.adsabs.harvard.edu/abs/2013MNRAS.435L..29S}
}

@ARTICLE{lsj+20,
       author = {{Lower}, Marcus E. and {Shannon}, Ryan M. and {Johnston}, Simon and {Bailes}, Matthew},
        title = "{Spectropolarimetric Properties of Swift J1818.0-1607: A 1.4 s Radio Magnetar}",
      journal = {\apjl},
         year = 2020,
        month = jun,
       volume = {896},
       number = {2},
          eid = {L37},
        pages = {L37},
          doi = {10.3847/2041-8213/ab9898},
archivePrefix = {arXiv},
       eprint = {2004.11522},
 primaryClass = {astro-ph.HE},
       adsurl = {https://ui.adsabs.harvard.edu/abs/2020ApJ...896L..37L}
}

@ARTICLE{ksj+21,
       author = {{Kirsten}, F. and {Snelders}, M.~P. and {Jenkins}, M. and {Nimmo}, K. and {van den Eijnden}, J. and {Hessels}, J.~W.~T. and {Gawro{\'n}ski}, M.~P. and {Yang}, J.},
        title = "{Detection of two bright radio bursts from magnetar SGR 1935 + 2154}",
      journal = {Nature Astronomy},
         year = 2021,
        month = apr,
       volume = {5},
        pages = {414-422},
          doi = {10.1038/s41550-020-01246-3},
archivePrefix = {arXiv},
       eprint = {2007.05101},
 primaryClass = {astro-ph.HE},
       adsurl = {https://ui.adsabs.harvard.edu/abs/2021NatAs...5..414K}
}

@ARTICLE{djw+18,
       author = {{Dai}, S. and {Johnston}, S. and {Weltevrede}, P. and {Kerr}, M. and {Burgay}, M. and {Esposito}, P. and {Israel}, G. and {Possenti}, A. and {Rea}, N. and {Sarkissian}, J.},
        title = "{Peculiar spin frequency and radio profile evolution of PSR J1119-6127 following magnetar-like X-ray bursts}",
      journal = {\mnras},
         year = 2018,
        month = nov,
       volume = {480},
       number = {3},
        pages = {3584-3594},
          doi = {10.1093/mnras/sty2063},
archivePrefix = {arXiv},
       eprint = {1806.05064},
 primaryClass = {astro-ph.HE},
       adsurl = {https://ui.adsabs.harvard.edu/abs/2018MNRAS.480.3584D}
}

@ARTICLE{rsl+22,
       author = {{Rajwade}, K.~M. and {Stappers}, B.~W. and {Lyne}, A.~G. and {Shaw}, B. and {Mickaliger}, M.~B. and {Liu}, K. and {Kramer}, M. and {Desvignes}, G. and {Karuppusamy}, R. and {Enoto}, T. and {G{\"u}ver}, T. and {Hu}, Chin-Ping and {Surnis}, M.~P.},
        title = "{Long term radio and X-ray evolution of the magnetar Swift J1818.0-1607}",
      journal = {\mnras},
         year = 2022,
        month = may,
       volume = {512},
       number = {2},
        pages = {1687-1695},
          doi = {10.1093/mnras/stac446},
archivePrefix = {arXiv},
       eprint = {2202.07548},
 primaryClass = {astro-ph.HE},
       adsurl = {https://ui.adsabs.harvard.edu/abs/2022MNRAS.512.1687R}
}

@ARTICLE{egk+20,
       author = {{Evans}, P.~A. and {Gropp}, J.~D. and {Kennea}, J.~A. and {Klingler}, N.~J. and {Laha}, S. and {Lien}, A.~Y. and {Page}, K.~L. and {Sakamoto}, T. and {Tohuvavohu}, A. and {Neil Gehrels Swift Observatory Team}},
        title = "{Swift-BAT trigger 960986: Swift detection of a new SGR Swift J1818.0-1607}",
      journal = {GRB Coordinates Network},
         year = 2020,
        month = mar,
       volume = {27373},
        pages = {1},
       adsurl = {https://ui.adsabs.harvard.edu/abs/2020GCN.27373....1E}
}

@ARTICLE{rsl+20,
       author = {{Rajwade}, Kaustubh and {Stappers}, Benjamin and {Lyne}, Andrew and {Mickaliger}, Mitchell B. and {Preston}, Lina Levin and {Keith}, Michael and {Weltevrede}, Patrick and {Kramer}, Michael and {van der Horst}, Alexander and {Kouveliotou}, Chryssa and {O'Connor}, Brendan},
        title = "{Confirmation of pulsed radio emission from Swift J1818.0-1607}",
      journal = {The Astronomer's Telegram},
         year = 2020,
        month = mar,
       volume = {13554},
        pages = {1},
       adsurl = {https://ui.adsabs.harvard.edu/abs/2020ATel13554....1R}
}

@ARTICLE{lkc+20,
       author = {{Liu}, Kuo and {Karuppusamy}, Ramesh and {Cognard}, Ismael and {Desvignes}, Gregory and {Kramer}, Michael and {Lyne}, Andrew and {Rajwade}, Kaustubh and {Stappers}, Ben and {Torne}, Pablo},
        title = "{Polarimetric detection of the magnetar Swift J1818.0-1607 from 4 to 22 GHz with the Effelsberg 100-m Telescope}",
      journal = {The Astronomer's Telegram},
         year = 2020,
        month = sep,
       volume = {13997},
        pages = {1},
       adsurl = {https://ui.adsabs.harvard.edu/abs/2020ATel13997....1L}
}

@ARTICLE{jb20,
       author = {{Joshi}, Bhal Chandra and {Bagchi}, Manjari},
        title = "{Detection of pulsed radio emission from bursting Magnetar Swift J1818.0-1607 below 750 MHz with the uGMRT}",
      journal = {The Astronomer's Telegram},
         year = 2020,
        month = mar,
       volume = {13580},
        pages = {1},
       adsurl = {https://ui.adsabs.harvard.edu/abs/2020ATel13580....1J}}

@ARTICLE{ibr+23,
       author = {{Ibrahim}, A.~Y. and {Borghese}, A. and {Rea}, N. and {Coti Zelati}, F. and {Parent}, E. and {Russell}, T.~D. and {Ascenzi}, S. and {Sathyaprakash}, R. and {G{\"o}tz}, D. and {Mereghetti}, S. and {Topinka}, M. and {Rigoselli}, M. and {Savchenko}, V. and {Campana}, S. and {Israel}, G.~L. and {Tiengo}, A. and {Perna}, R. and {Turolla}, R. and {Zane}, S. and {Esposito}, P. and {Rodr{\'\i}guez Castillo}, G.~A. and {Graber}, V. and {Possenti}, A. and {Dehman}, C. and {Ronchi}, M. and {Loru}, S.},
        title = "{Deep X-Ray and Radio Observations of the First Outburst of the Young Magnetar Swift J1818.0-1607}",
      journal = {\apj},
         year = 2023,
        month = jan,
       volume = {943},
       number = {1},
          eid = {20},
        pages = {20},
          doi = {10.3847/1538-4357/aca528},
archivePrefix = {arXiv},
       eprint = {2211.12391},
 primaryClass = {astro-ph.HE},
       adsurl = {https://ui.adsabs.harvard.edu/abs/2023ApJ...943...20I}
}

@ARTICLE{gak+17,
       author = {{Gupta}, Y. and {Ajithkumar}, B. and {Kale}, H.~S. and {Nayak}, S. and {Sabhapathy}, S. and {Sureshkumar}, S. and {Swami}, R.~V. and {Chengalur}, J.~N. and {Ghosh}, S.~K. and {Ishwara-Chandra}, C.~H. and {Joshi}, B.~C. and {Kanekar}, N. and {Lal}, D.~V. and {Roy}, S.},
        title = "{The upgraded GMRT: opening new windows on the radio Universe}",
      journal = {Current Science},
         year = 2017,
        month = aug,
       volume = {113},
       number = {4},
        pages = {707-714},
          doi = {10.18520/cs/v113/i04/707-714},
       adsurl = {https://ui.adsabs.harvard.edu/abs/2017CSci..113..707G}
}

@ARTICLE{ki+21,
       author = {{Kale}, Ruta and {Ishwara-Chandra}, C.~H.},
        title = "{CAPTURE: a continuum imaging pipeline for the uGMRT}",
      journal = {Experimental Astronomy},
         year = 2021,
        month = feb,
       volume = {51},
       number = {1},
        pages = {95-108},
          doi = {10.1007/s10686-020-09677-6},
archivePrefix = {arXiv},
       eprint = {2010.00196},
 primaryClass = {astro-ph.IM},
       adsurl = {https://ui.adsabs.harvard.edu/abs/2021ExA....51...95K}
}

@ARTICLE{casa22,
       author = {{CASA Team} and {Bean}, Ben and {Bhatnagar}, Sanjay and {Castro}, Sandra and {Donovan Meyer}, Jennifer and {Emonts}, Bjorn and {Garcia}, Enrique and {Garwood}, Robert and {Golap}, Kumar and {Gonzalez Villalba}, Justo and {Harris}, Pamela and {Hayashi}, Yohei and {Hoskins}, Josh and {Hsieh}, Mingyu and {Jagannathan}, Preshanth and {Kawasaki}, Wataru and {Keimpema}, Aard and {Kettenis}, Mark and {Lopez}, Jorge and {Marvil}, Joshua and {Masters}, Joseph and {McNichols}, Andrew and {Mehringer}, David and {Miel}, Renaud and {Moellenbrock}, George and {Montesino}, Federico and {Nakazato}, Takeshi and {Ott}, Juergen and {Petry}, Dirk and {Pokorny}, Martin and {Raba}, Ryan and {Rau}, Urvashi and {Schiebel}, Darrell and {Schweighart}, Neal and {Sekhar}, Srikrishna and {Shimada}, Kazuhiko and {Small}, Des and {Steeb}, Jan-Willem and {Sugimoto}, Kanako and {Suoranta}, Ville and {Tsutsumi}, Takahiro and {van Bemmel}, Ilse M. and {Verkouter}, Marjolein and {Wells}, Akeem and {Xiong}, Wei and {Szomoru}, Arpad and {Griffith}, Morgan and {Glendenning}, Brian and {Kern}, Jeff},
        title = "{CASA, the Common Astronomy Software Applications for Radio Astronomy}",
      journal = {\pasp},
         year = 2022,
        month = nov,
       volume = {134},
       number = {1041},
          eid = {114501},
        pages = {114501},
          doi = {10.1088/1538-3873/ac9642},
archivePrefix = {arXiv},
       eprint = {2210.02276},
 primaryClass = {astro-ph.IM},
       adsurl = {https://ui.adsabs.harvard.edu/abs/2022PASP..134k4501C}
}

@ARTICLE{ijm+17,
       author = {{Intema}, H.~T. and {Jagannathan}, P. and {Mooley}, K.~P. and {Frail}, D.~A.},
        title = "{The GMRT 150 MHz all-sky radio survey. First alternative data release TGSS ADR1}",
      journal = {\aap},
         year = 2017,
        month = feb,
       volume = {598},
          eid = {A78},
        pages = {A78},
          doi = {10.1051/0004-6361/201628536},
archivePrefix = {arXiv},
       eprint = {1603.04368},
 primaryClass = {astro-ph.CO},
       adsurl = {https://ui.adsabs.harvard.edu/abs/2017A&A...598A..78I}
}

@ARTICLE{hmt+21,
       author = {{Hale}, Catherine L. and {McConnell}, D. and {Thomson}, A.~J.~M. and {Lenc}, E. and {Heald}, G.~H. and {Hotan}, A.~W. and {Leung}, J.~K. and {Moss}, V.~A. and {Murphy}, T. and {Pritchard}, J. and {Sadler}, E.~M. and {Stewart}, A.~J. and {Whiting}, M.~T.},
        title = "{The Rapid ASKAP Continuum Survey Paper II: First Stokes I Source Catalogue Data Release}",
      journal = {\pasa},
         year = 2021,
        month = nov,
       volume = {38},
          eid = {e058},
        pages = {e058},
          doi = {10.1017/pasa.2021.47},
archivePrefix = {arXiv},
       eprint = {2109.00956},
 primaryClass = {astro-ph.GA},
       adsurl = {https://ui.adsabs.harvard.edu/abs/2021PASA...38...58H}
}

@ARTICLE{dgh+24,
       author = {{Duchesne}, S.~W. and {Grundy}, J.~A. and {Heald}, George H. and {Lenc}, Emil and {Leung}, James K. and {McConnell}, David and {Murphy}, Tara and {Pritchard}, Joshua and {Rose}, Kovi and {Thomson}, Alec J.~M. and {Wang}, Yuanming and {Wang}, Ziteng and {Whiting}, Matthew T.},
        title = "{The Rapid ASKAP Continuum Survey V: Cataloguing the sky at 1 367.5 MHz and the second data release of RACS-mid}",
      journal = {\pasa},
         year = 2024,
        month = jan,
       volume = {41},
          eid = {e003},
        pages = {e003},
          doi = {10.1017/pasa.2023.60},
archivePrefix = {arXiv},
       eprint = {2311.12369},
 primaryClass = {astro-ph.GA},
       adsurl = {https://ui.adsabs.harvard.edu/abs/2024PASA...41....3D}
}

@ARTICLE{SWF+24,
       author = {{Goedhart}, S. and {Cotton}, W.~D. and {Camilo}, F. and {Thompson}, M.~A. and {Umana}, G. and {Bietenholz}, M. and {Woudt}, P.~A. and {Anderson}, L.~D. and {Bordiu}, C. and {Buckley}, D.~A.~H. and {Buemi}, C.~S. and {Bufano}, F. and {Cavallaro}, F. and {Chen}, H. and {Chibueze}, J.~O. and {Egbo}, D. and {Frank}, B.~S. and {Hoare}, M.~G. and {Ingallinera}, A. and {Irabor}, T. and {Kraan-Korteweg}, R.~C. and {Kurapati}, S. and {Leto}, P. and {Loru}, S. and {Mutale}, M. and {Obonyo}, W.~O. and {Plavin}, A. and {Rajohnson}, S.~H.~A. and {Rigby}, A. and {Riggi}, S. and {Seidu}, M. and {Serra}, P. and {Smart}, B.~M. and {Stappers}, B.~W. and {Steyn}, N. and {Surnis}, M. and {Trigilio}, C. and {Williams}, G.~M. and {Abbott}, T.~D. and {Adam}, R.~M. and {Asad}, K.~M.~B. and {Baloyi}, T. and {Bauermeister}, E.~F. and {Bennet}, T.~G.~H. and {Bester}, H. and {Botha}, A.~G. and {Brederode}, L.~R.~S. and {Buchner}, S. and {Burger}, J.~P. and {Cheetham}, T. and {Cloete}, K. and {de Villiers}, M.~S. and {de Villiers}, D.~I.~L. and {du Toit}, L.~J. and {Esterhuyse}, S.~W.~P. and {Fanaroff}, B.~L. and {Fourie}, D.~J. and {Gamatham}, R.~R.~G. and {Gatsi}, T.~G. and {Geyer}, M. and {Gouws}, M. and {Gumede}, S.~C. and {Heywood}, I. and {Hokwana}, A. and {Hoosen}, S.~W. and {Horn}, D.~M. and {Horrell}, L.~M.~G. and {Hugo}, B.~V. and {Isaacson}, A.~I. and {J{\'o}zsa}, G.~I.~G. and {Jonas}, J.~L. and {Jordaan}, J.~D.~B.~L. and {Joubert}, A.~F. and {Julie}, R.~P.~M. and {Kapp}, F.~B. and {Kriek}, N. and {Kriel}, H. and {Krishnan}, V.~K. and {Kusel}, T.~W. and {Legodi}, L.~S. and {Lehmensiek}, R. and {Lord}, R.~T. and {Macfarlane}, P.~S. and {Magnus}, L.~G. and {Magozore}, C. and {Main}, J.~P.~L. and {Malan}, J.~A. and {Manley}, J.~R. and {Marais}, S.~J. and {Maree}, M.~D.~J. and {Martens}, A. and {Maruping}, P. and {McAlpine}, K. and {Merry}, B.~C. and {Mgodeli}, M. and {Millenaar}, R.~P. and {Mokone}, O.~J. and {Monama}, T.~E. and {New}, W.~S. and {Ngcebetsha}, B. and {Ngoasheng}, K.~J. and {Nicolson}, G.~D. and {Ockards}, M.~T. and {Oozeer}, N. and {Passmoor}, S.~S. and {Patel}, A.~A. and {Peens-Hough}, A. and {Perkins}, S.~J. and {Ramaila}, A.~J.~T. and {Ratcliffe}, S.~M. and {Renil}, R. and {Richter}, L.~L. and {Salie}, S. and {Sambu}, N. and {Schollar}, C.~T.~G. and {Schwardt}, L.~C. and {Schwartz}, R.~L. and {Serylak}, M. and {Siebrits}, R. and {Sirothia}, S.~K. and {Slabber}, M.~J. and {Smirnov}, O.~M. and {Tiplady}, A.~J. and {van Balla}, T.~J. and {van der Byl}, A. and {Van Tonder}, V. and {Venter}, A.~J. and {Venter}, M. and {Welz}, M.~G. and {Williams}, L.~P.},
        title = "{The SARAO MeerKAT 1.3 GHz Galactic Plane Survey}",
      journal = {\mnras},
         year = 2024,
        month = jun,
       volume = {531},
       number = {1},
        pages = {649-681},
          doi = {10.1093/mnras/stae1166},
archivePrefix = {arXiv},
       eprint = {2312.07275},
 primaryClass = {astro-ph.GA},
       adsurl = {https://ui.adsabs.harvard.edu/abs/2024MNRAS.531..649G}
}

@ARTICLE{blr01,
       author = {{Borkowski}, Kazimierz J. and {Lyerly}, William J. and {Reynolds}, Stephen P.},
        title = "{Supernova Remnants in the Sedov Expansion Phase: Thermal X-Ray Emission}",
      journal = {\apj},
         year = 2001,
        month = feb,
       volume = {548},
       number = {2},
        pages = {820-835},
          doi = {10.1086/319011},
archivePrefix = {arXiv},
       eprint = {astro-ph/0008066},
 primaryClass = {astro-ph},
       adsurl = {https://ui.adsabs.harvard.edu/abs/2001ApJ...548..820B}
}

@ARTICLE{ybm+19,
       author = {{Maan}, Yogesh and {Joshi}, Bhal Chandra and {Surnis}, Mayuresh P. and {Bagchi}, Manjari and {Manoharan}, P.~K.},
        title = "{Distinct Properties of the Radio Burst Emission from the Magnetar XTE J1810-197}",
      journal = {\apjl},
         year = 2019,
        month = sep,
       volume = {882},
       number = {1},
          eid = {L9},
        pages = {L9},
          doi = {10.3847/2041-8213/ab3a47},
archivePrefix = {arXiv},
       eprint = {1908.04304},
 primaryClass = {astro-ph.HE},
       adsurl = {https://ui.adsabs.harvard.edu/abs/2019ApJ...882L...9M}
}

@ARTICLE{rs75,
   author = {{Ruderman}, M.~A. and {Sutherland}, P.~G.},
    title = "{Theory of pulsars - Polar caps, sparks, and coherent microwave radiation}",
  journal = {\apj},
     year = 1975,
    month = feb,
   volume = 196,
    pages = {51-72},
      doi = {10.1086/153393},
   adsurl = {http://adsabs.harvard.edu/abs/1975ApJ...196...51R}
}

@article{mpd+2017,
  title={Post-outburst radio observations of the high magnetic field pulsar PSR J1119-6127},
  author={Majid, Walid A and Pearlman, Aaron B and Dobreva, Tatyana and Horiuchi, Shinji and Kocz, Jonathon and Lippuner, Jonas and Prince, Thomas A},
  journal={The Astrophysical Journal Letters},
  volume={834},
  number={1},
  pages={L2},
  year={2017},
  publisher={The American Astronomical Society}
}

@article{scholz2017,
  title={Spin-down evolution and radio disappearance of the magnetar PSR J1622--4950},
  author={Scholz, P and Camilo, F and Sarkissian, J and Reynolds, JE and Levin, Lina and Bailes, Matthew and Burgay, MARTA and Johnston, S and Kramer, Michael and Possenti, ANDREA},
  journal={The Astrophysical Journal},
  volume={841},
  number={2},
  pages={126},
  year={2017},
  publisher={The American Astronomical Society}
}

@ARTICLE{msj+22,
       author = {{Maan}, Yogesh and {Surnis}, Mayuresh P. and {Chandra Joshi}, Bhal and {Bagchi}, Manjari},
        title = "{Magnetar XTE J1810-197: Spectro-temporal Evolution of Average Radio Emission}",
      journal = {\apj},
         year = 2022,
        month = may,
       volume = {931},
       number = {1},
          eid = {67},
        pages = {67},
          doi = {10.3847/1538-4357/ac68f1},
archivePrefix = {arXiv},
       eprint = {2201.13006},
 primaryClass = {astro-ph.HE},
       adsurl = {https://ui.adsabs.harvard.edu/abs/2022ApJ...931...67M}
}

@software{carta,
       author = {{Comrie}, Angus and {Wang}, Kuo-Song and {Hsu}, Shou-Chieh and {Moraghan}, Anthony and {Harris}, Pamela and {Pang}, Qi and {Pi{\'n}ska}, Adrianna and {Chiang}, Cheng-Chin and {Chang}, Tien-Hao and {Hwang}, Yu-Hsuan and {Jan}, Hengtai and {Lin}, Ming-Yi and {Simmonds}, Rob},
        title = "{CARTA: The Cube Analysis and Rendering Tool for Astronomy}",
         year = 2021,
        month = jun,
          eid = {10.5281/zenodo.3377984},
          doi = {10.5281/zenodo.3377984},
      version = {2.0.0},
    publisher = {Zenodo},
       adsurl = {https://ui.adsabs.harvard.edu/abs/2021zndo...3377984C}
}

@ARTICLE{acf+25,
       author = {{Anderson}, L.~D. and {Camilo}, F. and {Faerber}, T. and {Bietenholz}, M. and {Bordiu}, C. and {Bufano}, F. and {Chibueze}, J.~O. and {Cotton}, W.~D. and {Ingallinera}, A. and {Loru}, S. and {Rigby}, A. and {Riggi}, S. and {Thompson}, M.~A. and {Trigilio}, C. and {Umana}, G. and {Williams}, G.~M.},
        title = "{Supernova remnant candidates discovered by the SARAO MeerKAT Galactic Plane Survey}",
      journal = {\aap},
         year = 2025,
        month = jan,
       volume = {693},
          eid = {A247},
        pages = {A247},
          doi = {10.1051/0004-6361/202451038},
archivePrefix = {arXiv},
       eprint = {2409.16607},
 primaryClass = {astro-ph.GA},
       adsurl = {https://ui.adsabs.harvard.edu/abs/2025A&A...693A.247A}
}

@ARTICLE{crj+08,
       author = {{Camilo}, F. and {Reynolds}, J. and {Johnston}, S. and {Halpern}, J.~P. and {Ransom}, S.~M.},
        title = "{The Magnetar 1E 1547.0-5408: Radio Spectrum, Polarimetry, and Timing}",
      journal = {\apj},
         year = 2008,
        month = may,
       volume = {679},
       number = {1},
        pages = {681-686},
          doi = {10.1086/587054},
archivePrefix = {arXiv},
       eprint = {0802.0494},
 primaryClass = {astro-ph},
       adsurl = {https://ui.adsabs.harvard.edu/abs/2008ApJ...679..681C}
}

@ARTICLE{tbb+22,
       author = {{Torne}, Pablo and {Bell}, Graham S. and {Bintley}, Dan and {Desvignes}, Gregory and {Berry}, David and {Dempsey}, Jessica T. and {Ho}, Paul T.~P. and {Parsons}, Harriet and {Eatough}, Ralph P. and {Karuppusamy}, Ramesh and {Kramer}, Michael and {Kramer}, Carsten and {Liu}, Kuo and {Paubert}, Gabriel and {Sanchez-Portal}, Miguel and {Schuster}, Karl F.},
        title = "{Submillimeter Pulsations from the Magnetar XTE J1810-197}",
      journal = {\apjl},
         year = 2022,
        month = feb,
       volume = {925},
       number = {2},
          eid = {L17},
        pages = {L17},
          doi = {10.3847/2041-8213/ac4caa},
archivePrefix = {arXiv},
       eprint = {2201.07820},
 primaryClass = {astro-ph.HE},
       adsurl = {https://ui.adsabs.harvard.edu/abs/2022ApJ...925L..17T}
}

@ARTICLE{lbl+25,
       author = {{Lewis}, Evan F. and {Blumer}, Harsha and {Lynch}, Ryan S. and {McLaughlin}, Maura A.},
        title = "{Multifrequency Radio Observations of the Magnetar Swift J1818.0─1607}",
      journal = {\apj},
         year = 2025,
        month = jul,
       volume = {988},
       number = {1},
          eid = {92},
        pages = {92},
          doi = {10.3847/1538-4357/ade14f},
archivePrefix = {arXiv},
       eprint = {2502.15200},
 primaryClass = {astro-ph.HE},
       adsurl = {https://ui.adsabs.harvard.edu/abs/2025ApJ...988...92L}
}

@ARTICLE{lls+18,
       author = {{Lyne}, Andrew and {Levin}, Lina and {Stappers}, Ben and {Mickaliger}, Mitch and {Desvignes}, Gregory and {Kramer}, Michael},
        title = "{Intense radio flare from the magnetar XTE J1810-197}",
      journal = {The Astronomer's Telegram},
         year = 2018,
        month = dec,
       volume = {12284},
        pages = {1},
       adsurl = {https://ui.adsabs.harvard.edu/abs/2018ATel12284....1L}
}

@ARTICLE{mnk+18,
       author = {{Mihara}, T. and {Negoro}, H. and {Kawai}, N. and {Nakajima}, M. and {Maruyama}, W. and {Sakamaki}, A. and {Aoki}, M. and {Kobayashi}, K. and {Nakahira}, S. and {Yatabe}, F. and {Takao}, Y. and {Matsuoka}, M. and {Sakamoto}, T. and {Serino}, M. and {Sugita}, S. and {Kawakubo}, Y. and {Hashimoto}, T. and {Yoshida}, A. and {Sugizaki}, M. and {Tachibana}, Y. and {Morita}, K. and {Oeda}, T. and {Shiraishi}, K. and {Ueno}, S. and {Tomida}, H. and {Ishikawa}, M. and {Sugawara}, Y. and {Isobe}, N. and {Shimomukai}, R. and {Midooka}, T. and {Ueda}, Y. and {Tanimoto}, A. and {Morita}, T. and {Yamada}, S. and {Ogawa}, S. and {Tsuboi}, Y. and {Iwakiri}, W. and {Sasaki}, R. and {Kawai}, H. and {Sato}, T. and {Tsunemi}, H. and {Yoneyama}, T. and {Asakura}, K. and {Ide}, S. and {Yamauchi}, M. and {Hidaka}, K. and {Iwahori}, S. and {Kurihara}, Y. and {Kawamuro}, T. and {Yamaoka}, K. and {Shidatsu}, M.},
        title = "{MAXI/GSC detection of the magnetar XTE J1810-197}",
      journal = {The Astronomer's Telegram},
         year = 2018,
        month = dec,
       volume = {12291},
        pages = {1},
       adsurl = {https://ui.adsabs.harvard.edu/abs/2018ATel12291....1M}
}

%%%%%%%%%%%%%%%%%%%%%%%%%%%%%%%%%%%%%%%%%%%%%%%%%%

%%%%%%%%%%%%%%%%% APPENDICES %%%%%%%%%%%%%%%%%%%%%

\appendix

\section{Contour Extraction and Flux Estimation}

To estimate the integrated flux density of the diffuse radio emission reported by \citet{ibr+23}, we extracted the published contour boundaries using an image-processing pipeline developed with the OpenCV library. Since the original FITS image was unavailable, the published contour map was used as the basis for the analysis. The image was converted to grayscale, smoothed using a Gaussian filter, and processed using the Canny edge-detection algorithm. A region-of-interest mask was applied to isolate the diffuse emission, followed by contour extraction using the OpenCV \texttt{findContours} routine.

\begin{figure}
    \centering
    \includegraphics[width=\linewidth]{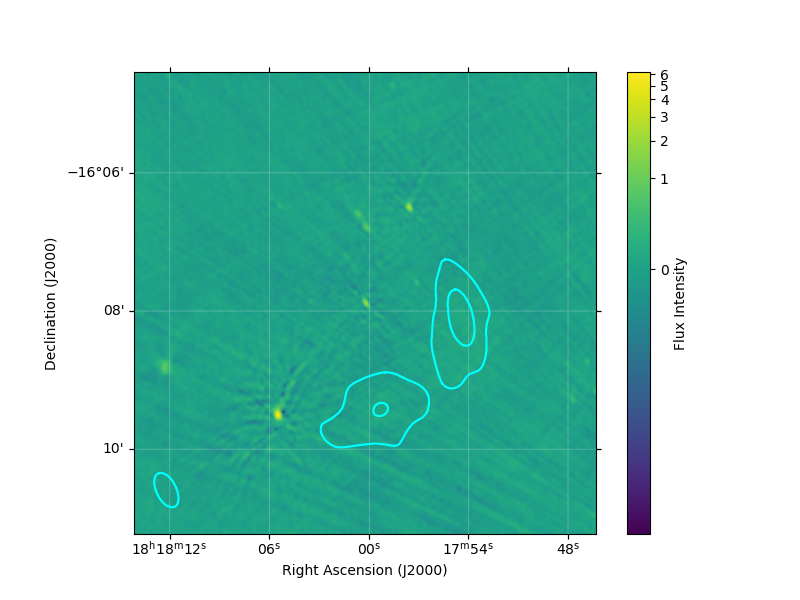}
    \caption{Final alignment of the extracted contours with our radio image, showing the diffuse emission regions used for the flux-density estimation.}
    \label{fig:contour_overlay}
\end{figure}

The extracted contours were registered onto our FITS image using an affine transformation. Several compact sources identifiable in both images were used as reference points to determine the transformation matrix. The transformation was applied to each contour using the OpenCV \texttt{transform} function. The resulting contours were exported as DS9 regions and overlaid on our radio image to verify their alignment. The final aligned contours are shown in Fig.~\ref{fig:contour_overlay}.

The area enclosed by each contour was calculated using the OpenCV \texttt{contourArea} function. The synthesized beam shown in the published image was extracted separately, and its area was measured using the same procedure. Since both areas were measured in pixel units, the number of synthesized beams enclosed by each contour was calculated as

\begin{equation}
N_{\rm beam} =
\frac{A_{\rm contour}}{A_{\rm beam}}.
\end{equation}

Assuming that the surface brightness within the outermost contour is approximately equal to the corresponding contour level, the integrated flux density was estimated as

\begin{equation}
S_{\rm int} =
I_{\rm contour}N_{\rm beam},
\end{equation}

where $I_{\rm contour}$ is the contour intensity in
$\mathrm{mJy\,beam^{-1}}$. The estimated flux densities of the two diffuse regions are $13.81$ and $10.37$ mJy, respectively, giving a total flux density of

\begin{equation}
S_{3000} = 24.18~\mathrm{mJy}.
\end{equation}

Assuming a power-law spectrum, $S_{\nu}\propto\nu^{\alpha}$, with $\alpha=-0.5$, the expected flux density at frequency $\nu$ is

\begin{equation}
S_{\nu}=S_{3000}
\left(\frac{\nu}{3000~\mathrm{MHz}}\right)^{-0.5}.
\end{equation}

This gives expected integrated flux densities of $52.23$ mJy at $643$ MHz and $66.22$ mJy at $400$ MHz.

\begin{table}[H]
\centering
\caption{Comparison between the expected flux densities extrapolated from the published 3 GHz image and the measured integrated flux densities in our uGMRT images.}
\label{tab:flux_comparison}
\begin{tabular}{lcc}
\hline
Quantity & Band 4 (643 MHz) & Band 3 (400 MHz) \\
\hline
Expected flux density & $52.23$ mJy & $66.22$ mJy \\
Measured flux density & $19.62$ mJy & $25.18$ mJy \\
Point-source contribution & $1.41 \pm 0.15$ mJy & -- \\
Diffuse flux density & $18.21$ mJy & $25.18$ mJy \\
RMS ($\sigma$) & $3.14$ mJy & $10.68$ mJy \\
Expected / RMS & $16.63$ & $6.20$ \\
Expected / measured diffuse flux & $2.86$ & $2.63$ \\
\hline
\end{tabular}
\end{table}

At $643$ MHz, the integrated flux densities measured within the two regions are $6.45$ and $13.17$ mJy, respectively, giving a total of $19.62$ mJy. A compact point source within Region~B contributes $1.41\pm0.15$ mJy, as measured using the CASA \texttt{imfit} task. After subtracting this contribution, the estimated diffuse emission is $18.21$ mJy.

At $400$ MHz, the corresponding integrated flux densities are $13.50$ and $11.68$ mJy, giving a total of $25.18$ mJy. No obvious compact point source was identified within these regions at this frequency.

The expected-to-measured flux-density ratios are therefore $2.86$ and $2.63$ at $643$ and $400$ MHz, respectively. The expected flux densities correspond to $16.63\sigma$ and $6.20\sigma$ when compared with the image RMS values of $3.14$ and $10.68$ mJy at $643$ and $400$ MHz, respectively. The RMS values for the extended regions were estimated from the CARTA-derived RMS using the $\sqrt{N}$ correction, where $N$ is the number of pixels within the corresponding extended region.

The lower measured flux densities relative to the extrapolated values may indicate that a significant fraction of the diffuse emission is resolved out or otherwise missed by the interferometric observations. However, the comparison should be treated with caution because the flux density estimated from the published contour map relies on the assumption of approximately uniform surface brightness within the outermost contour and on a fixed spectral index of $\alpha=-0.5$.

\section{Images with Different Values of the Robust Parameter}

We present the combined band 3 (Figures B1 and B2) and band 4 (Figures B3 and B4) images with different values of the robust parameter (0, 0.5, 1, and 2).  

\begin{figure*}
\centering
\begin{tabular}{cc}
\includegraphics[width=0.5\linewidth]{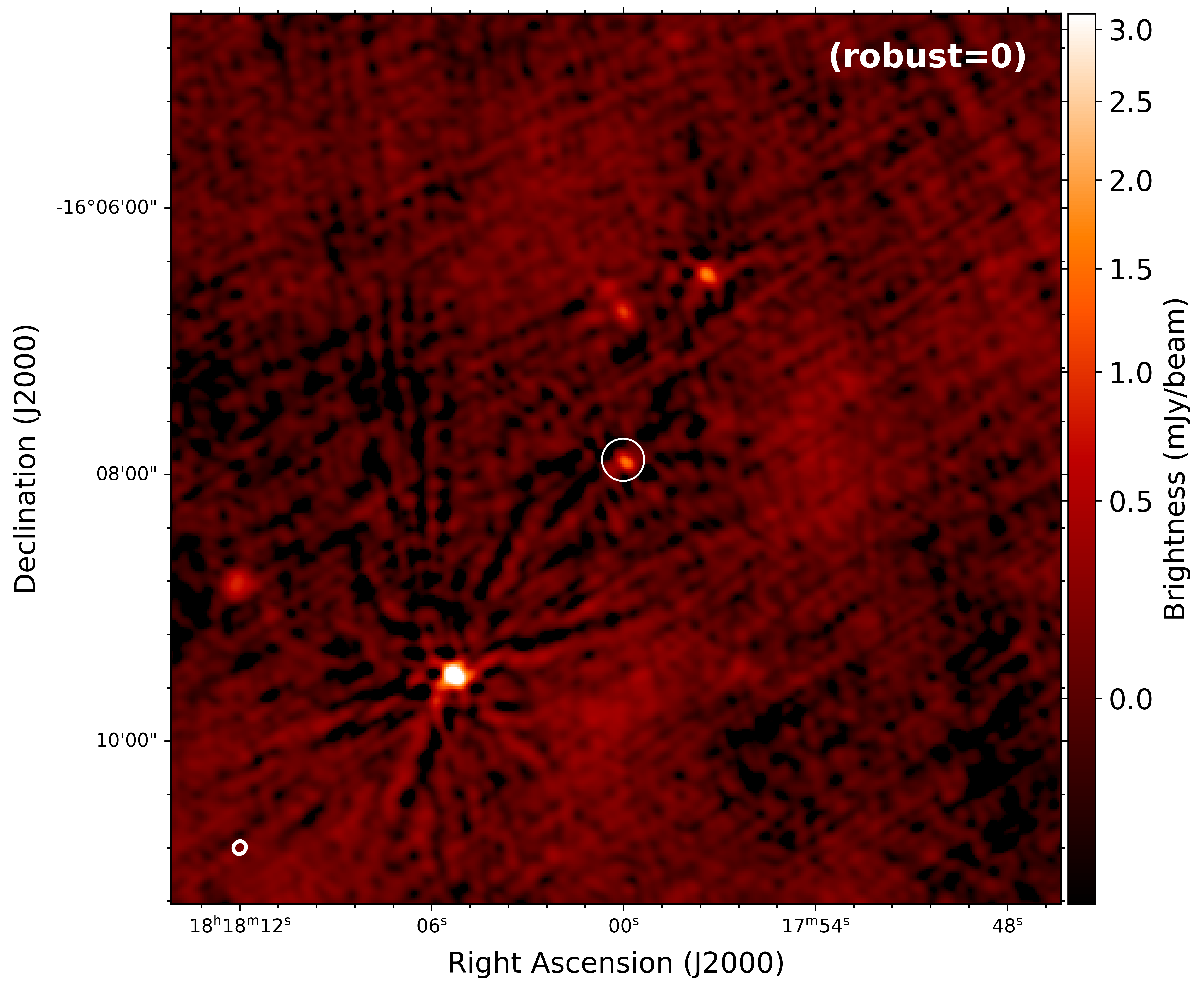} &
\includegraphics[width=0.5\linewidth]{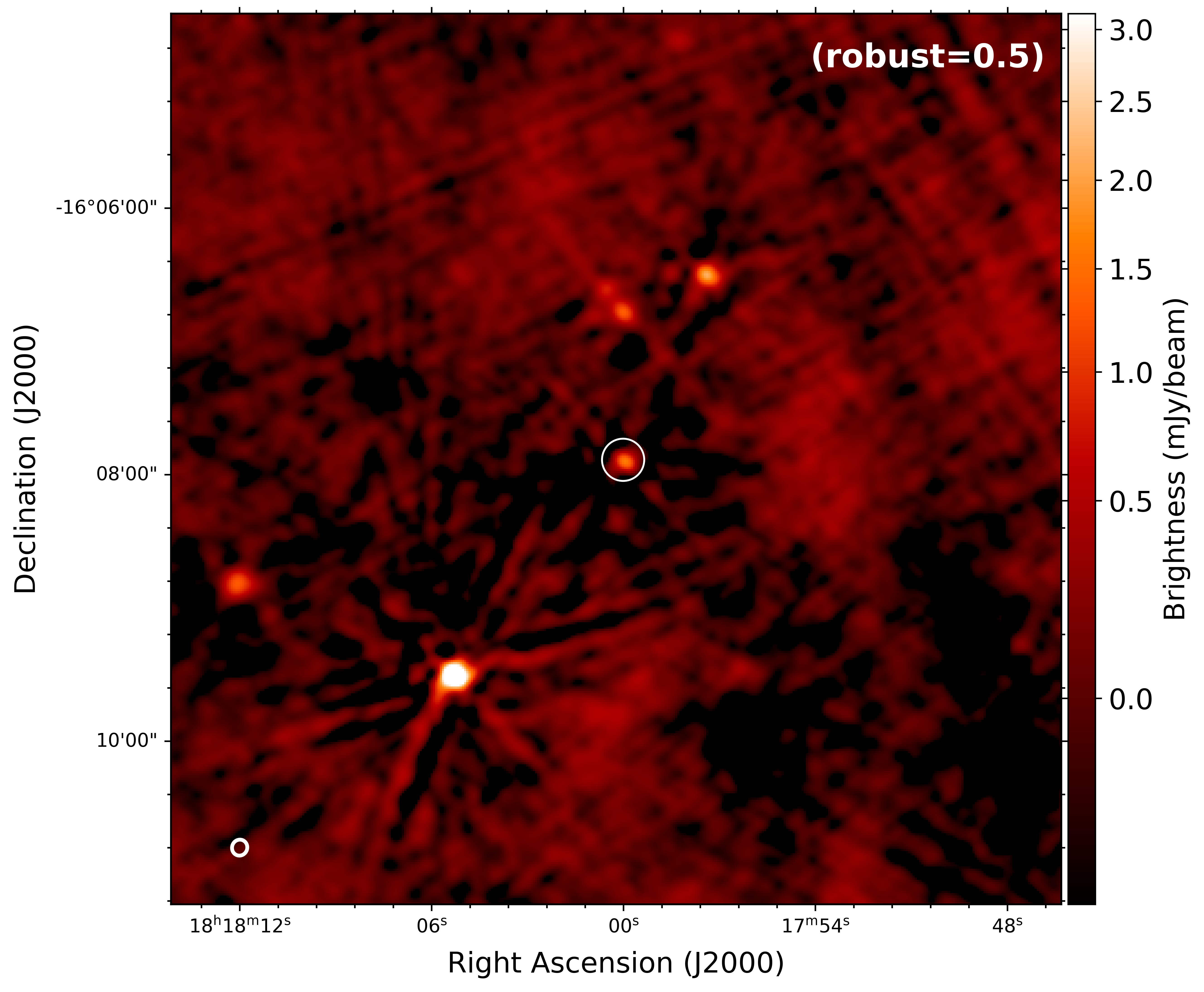} \\

\end{tabular}
\caption{Combined band 3 images with robust parameter value of 0 (left panel) and 0.5 (right panel).}
\end{figure*}

\begin{figure*}
\centering
\begin{tabular}{cc}

\includegraphics[width=0.5\linewidth]{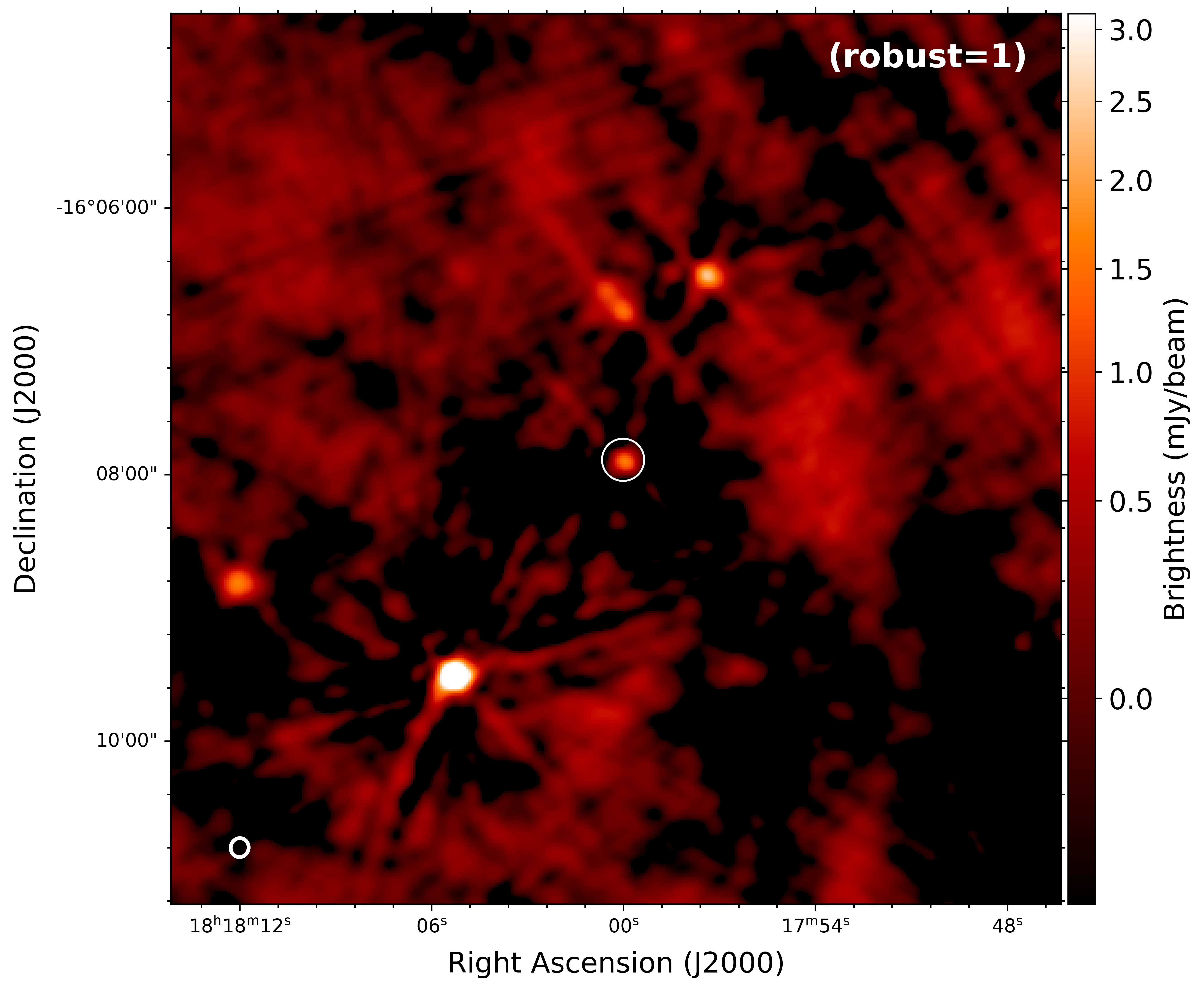} &
\includegraphics[width=0.5\linewidth]{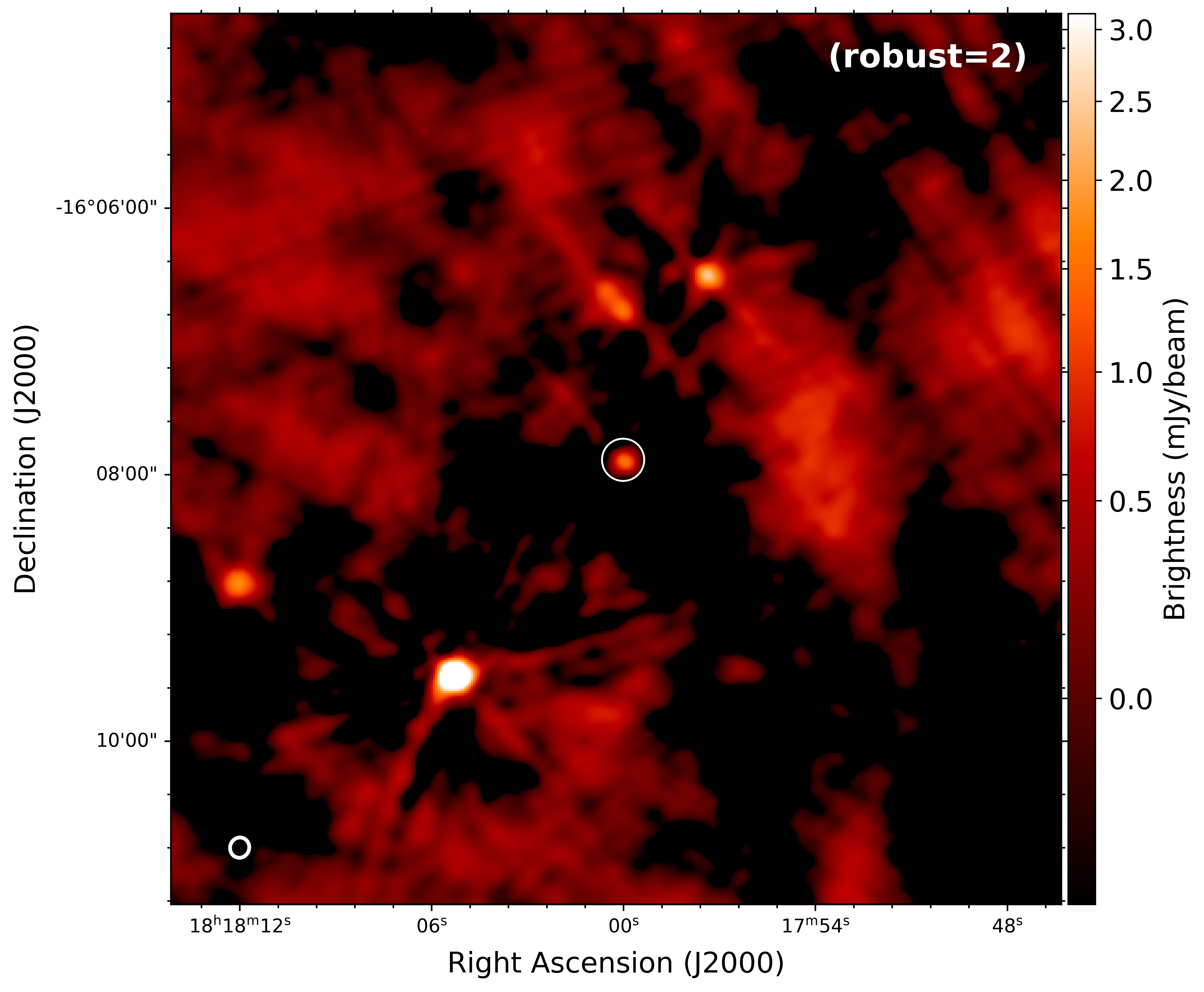} \\

\end{tabular}
\caption{Combined band 3 images with robust parameter value of 1 (left panel) and 2 (right panel).}
\end{figure*}

\begin{figure*}
\centering
\begin{tabular}{cc}
\includegraphics[width=0.5\linewidth]{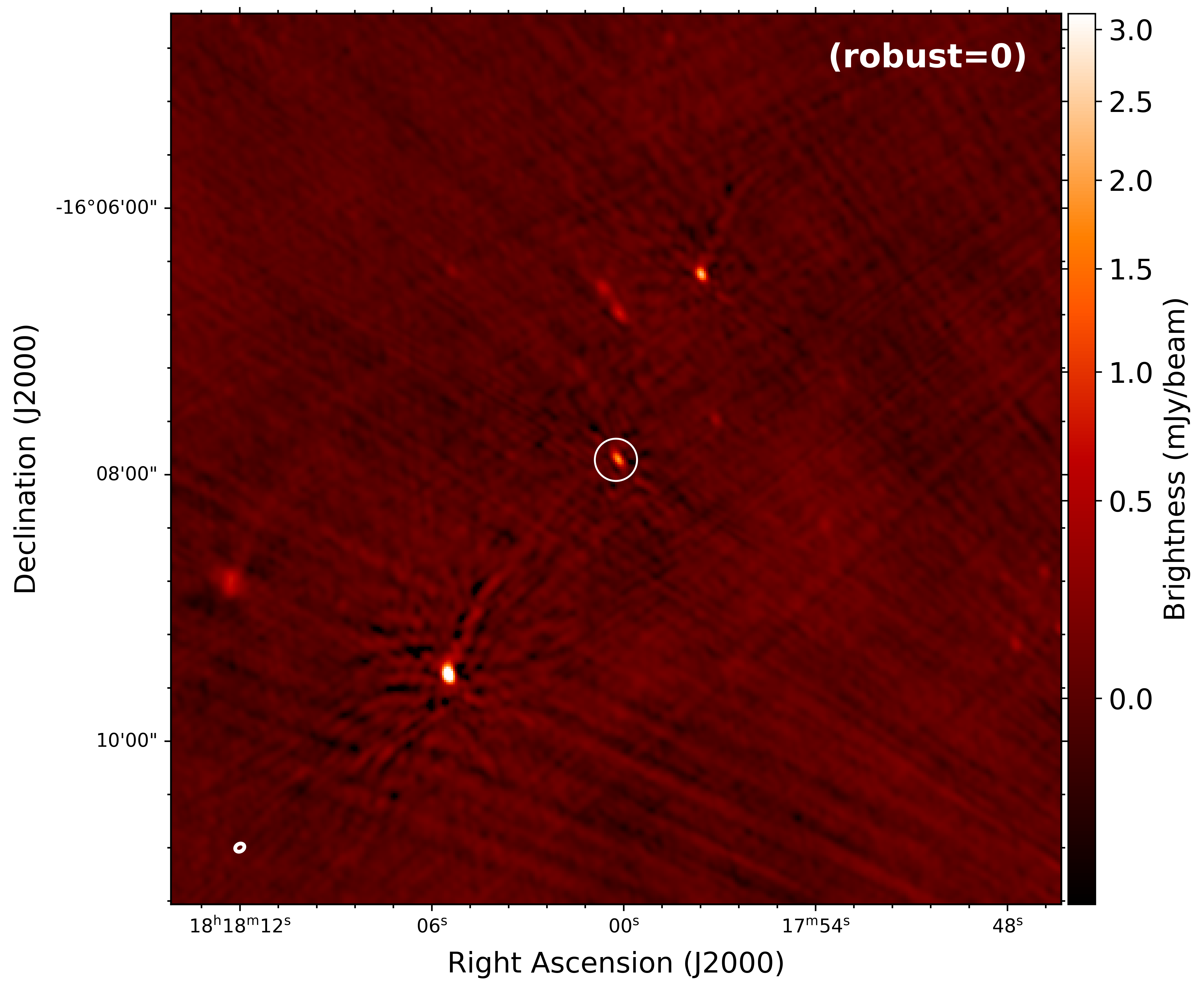} &
\includegraphics[width=0.5\linewidth]{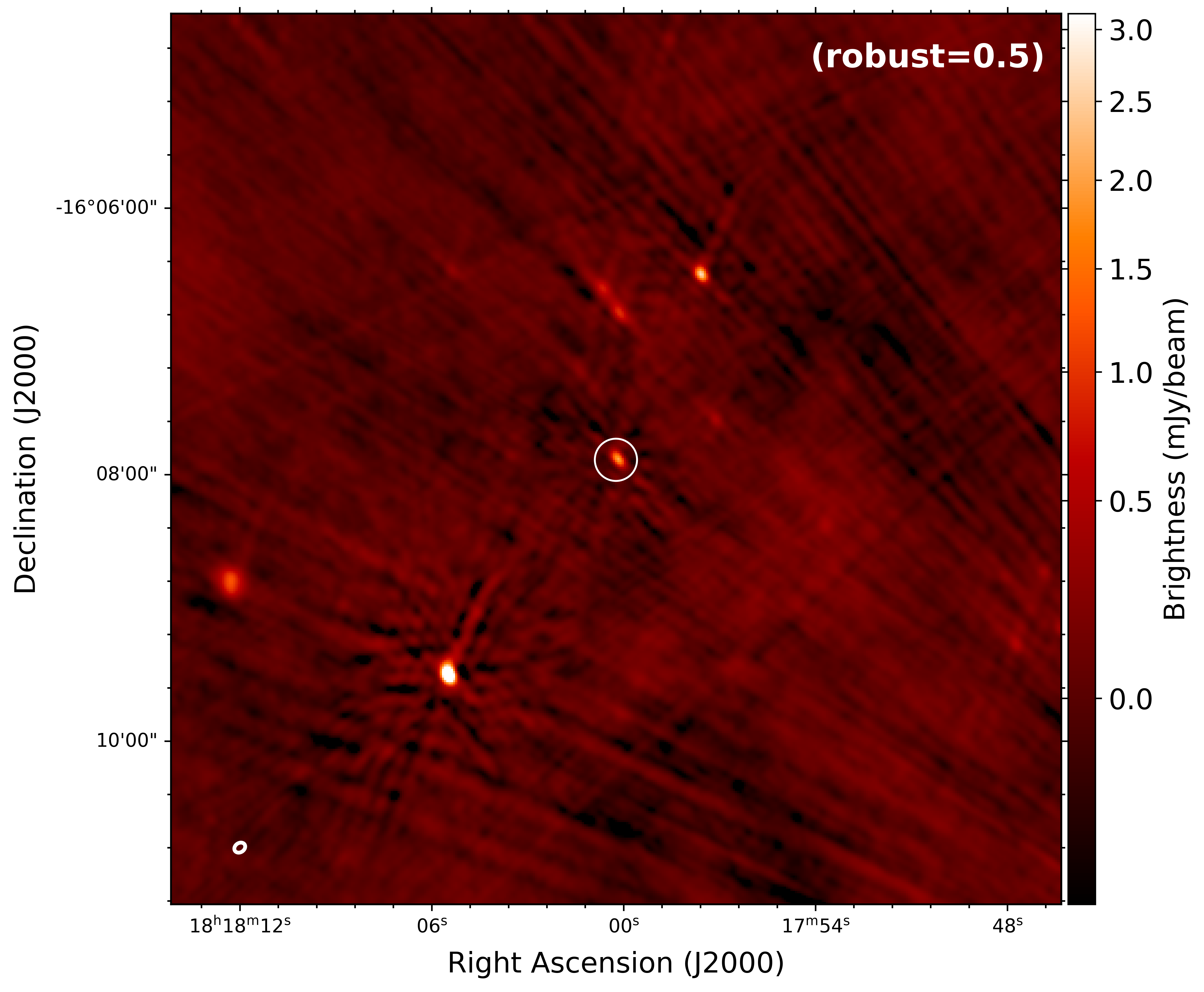} \\

\end{tabular}
\caption{Combined band 4 images with robust parameter value of 0 (left panel) and 0.5 (right panel).}
\end{figure*}

\begin{figure*}
\centering
\begin{tabular}{cc}

\includegraphics[width=0.5\linewidth]{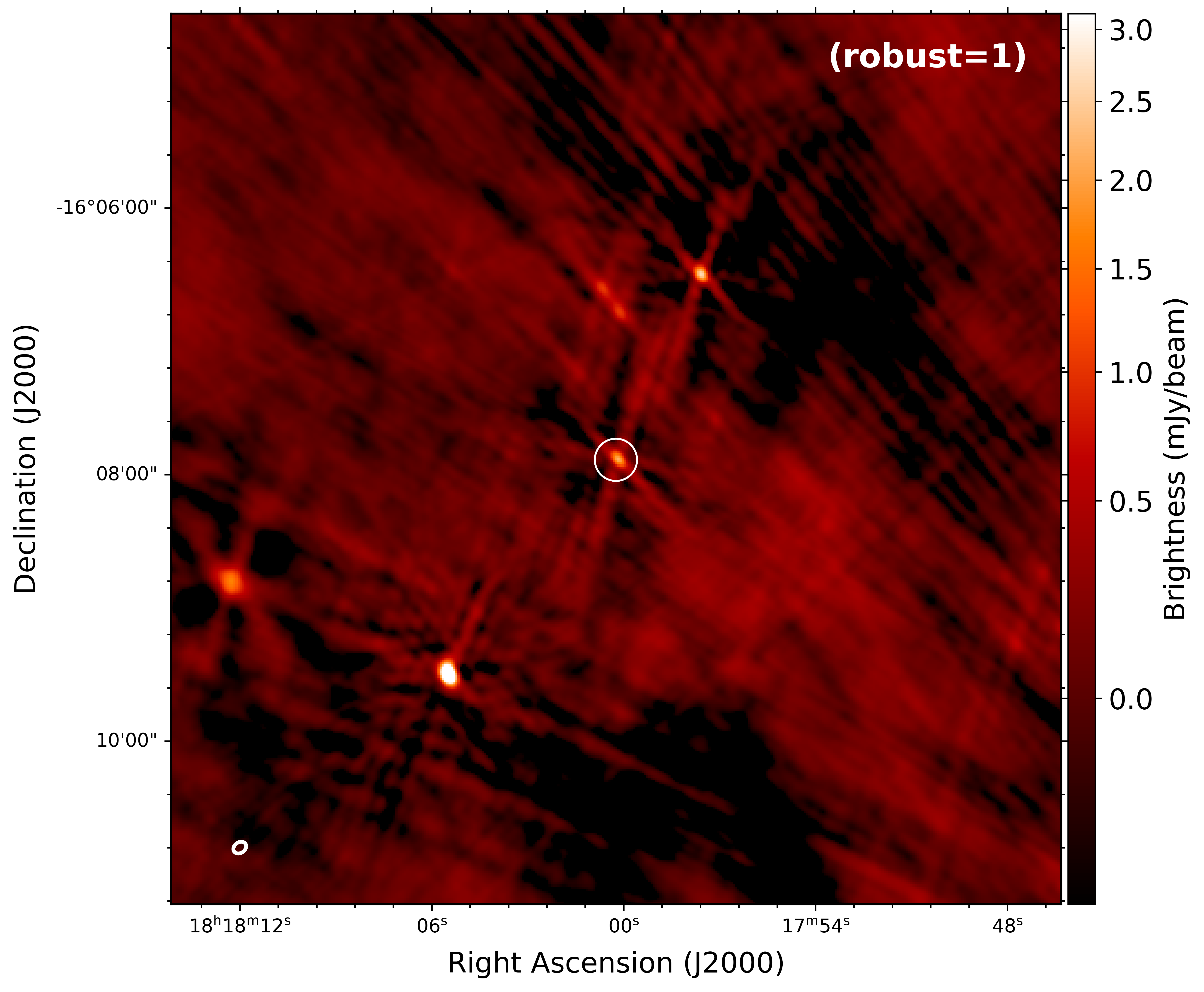} &
\includegraphics[width=0.5\linewidth]{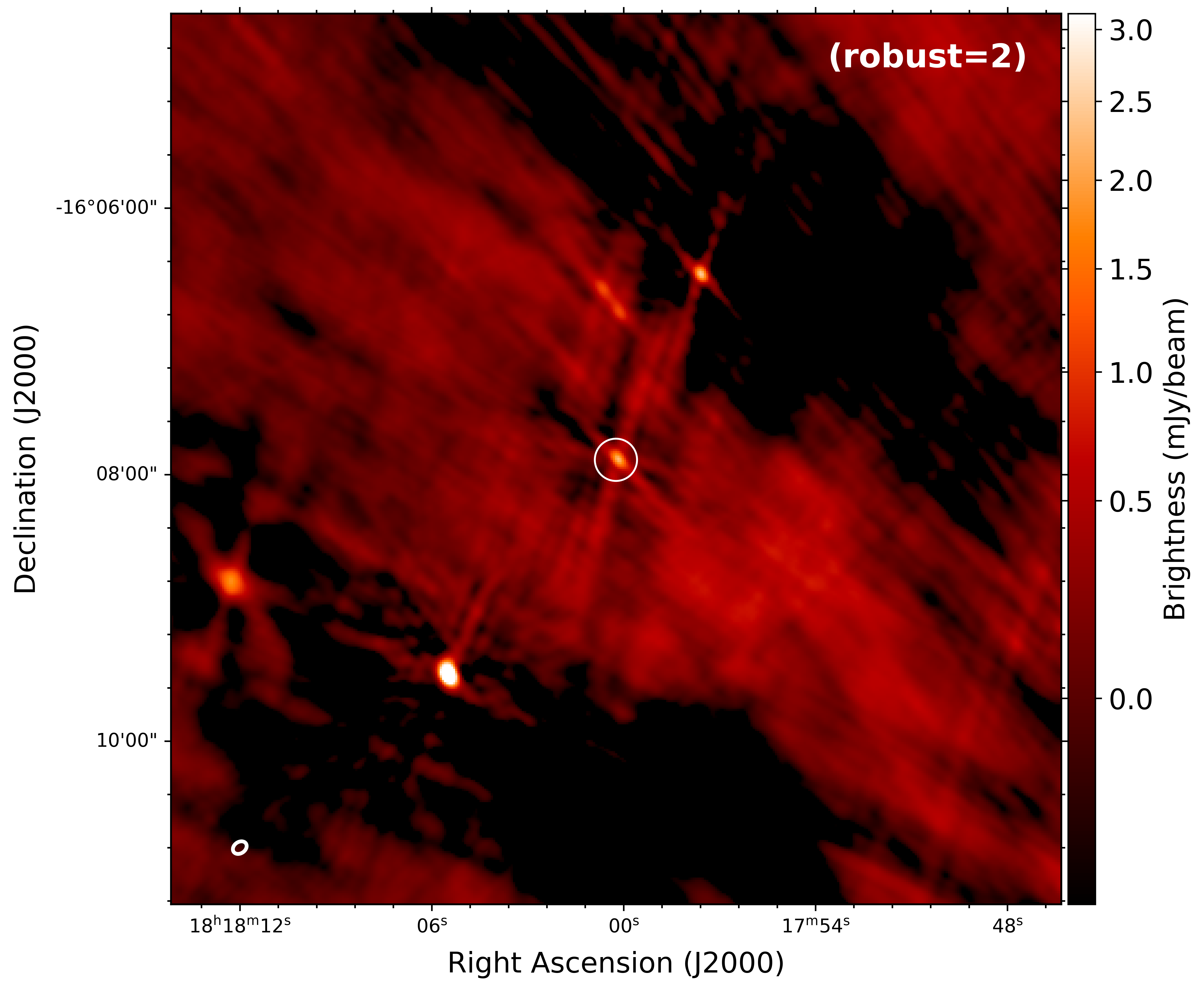} \\

\end{tabular}
\caption{Combined band 4 images with robust parameter value of 1 (left panel) and 2 (right panel).}
\end{figure*}

%%%%%%%%%%%%%%%%%%%%%%%%%%%%%%%%%%%%%%%%%%%%%%%%%%

% Don't change these lines
\bsp	% typesetting comment
\label{lastpage}
\end{document}